\documentclass[lettersize,journal]{IEEEtran}
\usepackage{amsmath,amsfonts}
\pdfoutput=1
\usepackage{algorithmic}
\usepackage{algorithm}
\usepackage{array}
\usepackage[caption=false,font=normalsize,labelfont=sf,textfont=sf]{subfig}
\usepackage{textcomp}
\usepackage{stfloats}
\usepackage{url}
\usepackage{verbatim}
\usepackage{graphicx}
\usepackage{cite}
\usepackage{xcolor}
\definecolor{dark-green}{HTML}{006400}
\definecolor{dark-blue}{HTML}{1976D2}
\definecolor{dark-purple}{HTML}{8d4bbb}
\definecolor{dark-red}{HTML}{D63C3C}
\definecolor{pink}{HTML}{fae6e9}
\definecolor{n1}{HTML}{ff9999}
\definecolor{n2}{HTML}{FFCC99}
\definecolor{n3}{HTML}{FFFF99}
\usepackage{hyperref}
\usepackage{hyperref}
\usepackage{colortbl}
\usepackage{threeparttable}
\usepackage{makecell}
\usepackage{multirow}
\usepackage{bbding}
\usepackage{pifont}

\hypersetup{
    colorlinks=true,
    linkcolor=blue,
    filecolor=magenta,      
    urlcolor=blue,
    citecolor=dark-blue,
}

\newcommand{\lyb}[1]{\textcolor{black}{#1}}

\newcommand{\bill}[1]{\textcolor{black}{#1}}

\begin{document}



\title{AccLLM: Accelerating Long-Context LLM Inference Via Algorithm-Hardware Co-Design}

\author{Yanbiao Liang, Huihong Shi, Haikuo Shao, and Zhongfeng Wang,~\IEEEmembership{Fellow,~IEEE}
\thanks{This work was supported by the National Key R\&D Program of China under Grant 2022YFB4400600.}
\thanks{Yanbiao Liang and Huihong Shi contributed equally to this work. Yanbiao Liang, Huihong Shi, and Haikuo Shao are with the School of Electronic Science and Engineering, Nanjing University, Nanjing, China (e-mail: \{{ybliang, shihh, hkshao}\}@smail.nju.edu.cn).}
\thanks{Zhongfeng Wang is with the School of Electronic Science and Engineering, Nanjing University, and the School of Integrated Circuits, Sun Yat-sen University (email: zfwang@nju.edu.cn).}
\thanks{Correspondence should be addressed to Zhongfeng Wang.}}




\maketitle

\begin{abstract}
Recently, large language models (LLMs) have achieved huge success in the natural language processing (NLP) field, driving a growing demand to extend their deployment from the cloud to edge devices. However, deploying LLMs on resource-constrained edge devices poses significant challenges, including (1) intensive computations and huge model sizes, (2) great memory and bandwidth demands introduced by the autoregressive generation process, and (3) limited scalability for handling long sequences.
To address these challenges, we propose AccLLM, a comprehensive acceleration framework that enables efficient and fast long-context LLM inference through algorithm and hardware co-design. At the algorithmic level, we integrate (1) pruning, (2) $\Lambda$-shaped attention, and (3) an innovative W$2$A$8$KV$4$ ($2$-bit weights, $8$-bit activations, and $4$-bit KV cache) quantization scheme, thus effectively reducing memory and bandwidth requirements while facilitating LLMs' long-sequence generation.
At the hardware level, we design a dedicated FPGA-based accelerator with a reconfigurable computing engine to effectively and flexibly accommodate diverse operations arising from our compression algorithm, thereby fully translating the algorithmic innovations into tangible hardware efficiency.
We validate AccLLM on the Xilinx Alveo U280 FPGA, demonstrating a \lyb{$\uparrow$$\mathbf{4.07}\times$} energy efficiency and a \lyb{$\uparrow$$\mathbf{2.98}\times$} throughput compared to the state-of-the-art work FlightLLM.
\end{abstract}


\section{Introduction}
\label{sec:intro}
Large language models \cite{Touvron2023LLaMAOA, Achiam2023GPT4TR, Zhang2022OPTOP, Touvron2023Llama2O} (LLMs) have revolutionized natural language processing (NLP) with their outstanding capabilities, enabling a wide range of applications \cite{Naveed2023ACO}, including code generation \cite{Chen2021EvaluatingLL}, document summarization \cite{Zhang2023BenchmarkingLL}, chatbots \cite{Achiam2023GPT4TR}, and question answering \cite{Kamalloo2023EvaluatingOQ}. This impressive potential has driven growing interest in extending LLMs' deploying beyond traditional cloud-based platforms to edge devices, such as smart vehicles, robots, and embedded systems \cite{Xu2024LlamaFAE, Huang2024EdgeLLMAH, Friha2024LLMBasedEI}. 
However, mainstream works have merely focused on optimizing and accelerating LLMs on GPUs \cite{Aminabadi2022DeepSpeedIE, Dao2022FlashAttentionFA} with powerful resource capacity, making them unsuitable for resource-constrained edge scenarios \cite{Kim2023FullSO, Pope2022EfficientlyST}.

\begin{figure}[!t]
\centering
\setlength{\abovecaptionskip}{0.1cm}
\includegraphics[width=\columnwidth]{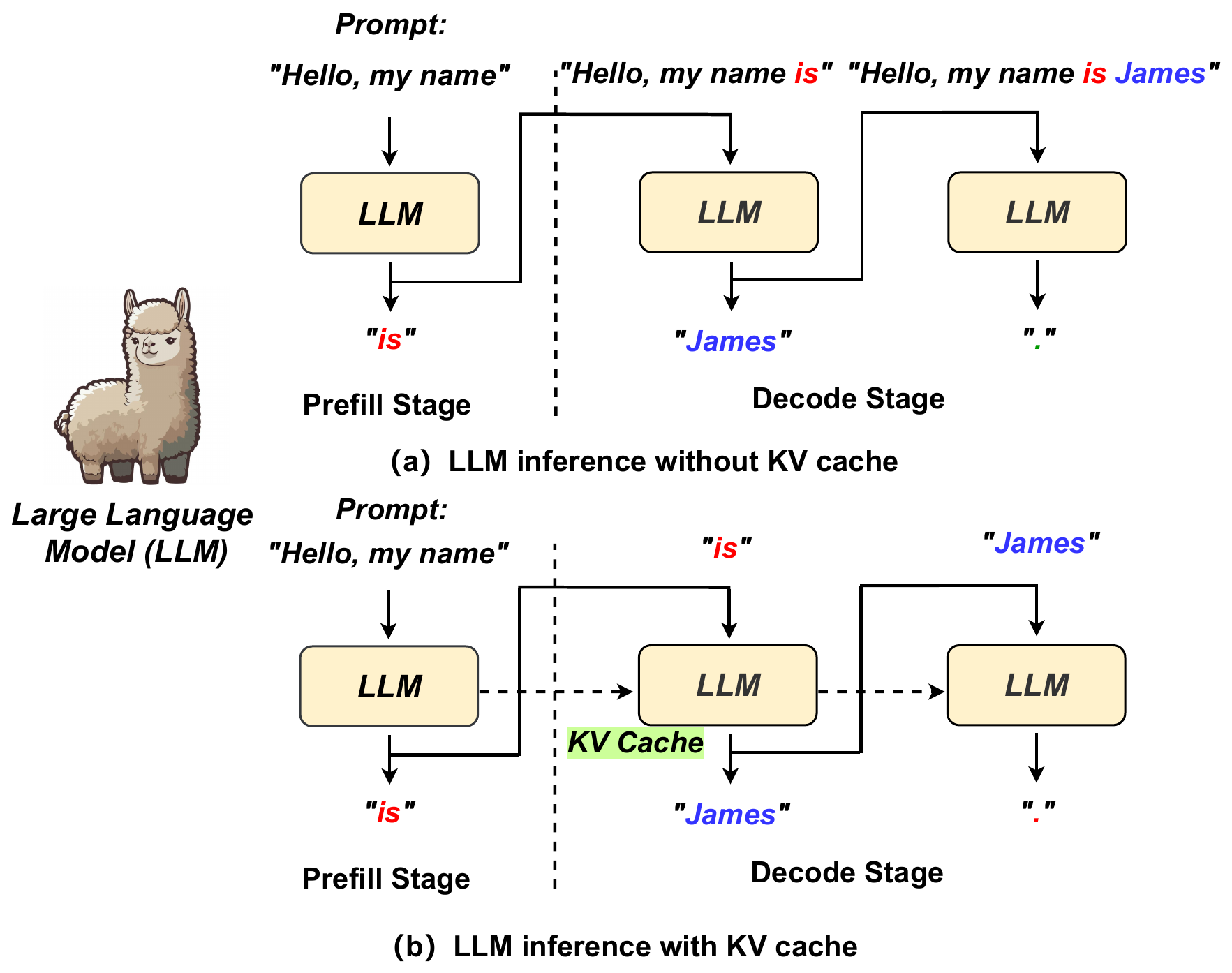}
\caption{The inference pipelines of LLMs (a) without KV cache and (b) with KV cache.}
\label{fig:LLMFlow}\vspace{-1.5em}
\end{figure}

To facilitate the widespread deployment of LLMs on edge devices, significant efforts \cite{Zeng2024FlightLLMEL, Huang2024EdgeLLMAH} have focused on developing hardware accelerators tailored for LLMs, with a particular emphasis on FPGAs due to their efficiency and reconfigurability. However, these approaches still face substantial design challenges.
\textit{\underline{\textbf{First}}, LLMs impose substantial computations and extremely large model sizes.} 
For example, a widely adopted LLM, Llama-2-7B \cite{Touvron2023Llama2O}, consists of $7$ billion parameters, requiring approximately $14$GB of memory in FP16 format and about $12$ TFLOPS of computation to perform a single inference with $512$ input tokens and $512$ output tokens. 
To tackle these issues, model compression techniques such as pruning \cite{Frantar2023SparseGPTML, Sun2023ASA} and quantization \cite{Lin2024QServeWQ, Xiao2022SmoothQuantAA, lin2024awq} have been proposed.
For example, SpareGPT \cite{Frantar2023SparseGPTML} uses pruning to enhance LLMs' efficiency at the architectural level by removing approximately ${50}$\% of unimportant components.
{In contrast, AWQ \cite{lin2024awq} proposes activation-aware weight quantization to quantize weights of LLMs to lower-bit integers, thus enhancing hardware efficiency at the operator level.}
Despite the promising results of these compression techniques, compressed LLMs remain too large for execution on edge devices with extremely constrained hardware resources, underscoring the need for more aggressive and effective model compression strategies.

\textit{\underline{\textbf{Second}}, the decode stage of LLMs exhibits substantial memory overhead and bandwidth demands.} As shown in Fig. \ref{fig:LLMFlow} (a), LLMs operate through two distinct stages \cite{Agrawal2023SARATHIEL}: the prefill and decode stages, each characterized by unique operation types and computational patterns.
Specifically, in the prefill stage, LLMs process all tokens in the input prompt in parallel to generate \bill{the first output token}. This stage is dominated by matrix-matrix (MM) multiplications, making it highly computationally intensive \cite{Zhao2024PrepackingAS}.
In contrast, the decode stage predicts output tokens sequentially, using both the prefill context and previously generated tokens. 
The autoregressive process in the decode stage involves repeated computations for each newly generated token, introducing redundancy. 
To mitigate this, as depicted in Fig. \ref{fig:LLMFlow} (b), caching mechanisms (i.e., Key-Value (KV) cache \cite{Ge2023ModelTY}) are proposed to store previous KV states and avoid redundant computations. However, this comes at the cost of increased memory overhead \cite{Xiao2023EfficientSL}, highlighting the need for efficient cache compression methods, particularly given the limited memory resources available on edge devices.
Moreover, the use of KV caches makes the decode stage primarily dependent on vector-matrix (VM) multiplications, rendering it memory-bound \cite{Kwon2023EfficientMM}.
Unfortunately, existing works are mainly dedicated to accelerating the prefill stage and cannot be effectively adapted to support the decode stage of LLMs, resulting in under-utilization of computing engines due to substantial memory and bandwidth demands associated with the decode stage \cite{Qin2023FACTFC, Ham2021ELSAHC, Lu2021SangerAC}. 



\begin{figure*}[!t]
\centering
\setlength{\abovecaptionskip}{0.1cm}
\includegraphics[scale=.3]{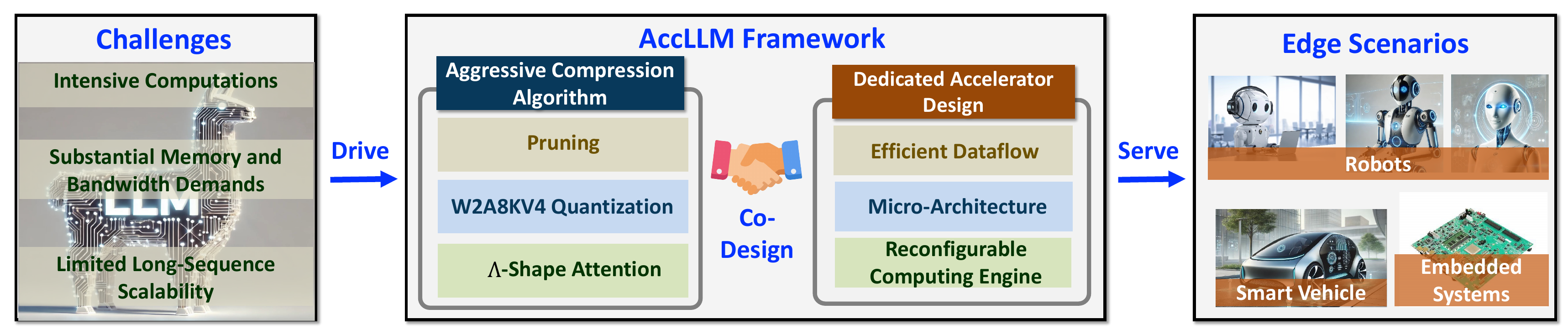}
\caption{Driven by the (1) intensive computations and large model sizes of LLMs, (2) substantial memory overhead and bandwidth demands introduced by the autoregressive generation process, and (3) limited scalability for handling long sequences, we propose a comprehensive LLM acceleration framework dubbed \textbf{AccLLM}, which incorporates (1) an aggressive compression algorithm and (2) a dedicated accelerator, thus facilitating extensive real-world applications of LLMs on edge devices. }
\label{fig:Overview}\vspace{-1.5em}
\end{figure*}

\textit{\underline{\textbf{Third}}, LLMs are generally required to handle long sequences but face significant challenges in terms of both performance and memory requirements.}
For example, chatbots need to process and maintain coherence over extended interactions, such as day-long conversations \cite{Achiam2023GPT4TR}. However, extending LLMs beyond their pre-trained sequence lengths often results in performance degradation \cite{Press2021TrainST}.
Additionally, the memory requirements for the KV cache during the autoregressive decode stage increase with sequence length \cite{Xiao2023EfficientSL}, further exacerbating the challenge of handling long sequences efficiently.
A promising solution to enhance LLMs' long-sequence scalability is $\Lambda$-shaped attention \cite{Xiao2023EfficientSL, han2024lm}, which combines global attention (targeting important initial tokens) with window attention (focusing on the most recent tokens). Therefore, effectively accelerating $\Lambda$-shaped attention and integrating it with other compression and optimization techniques is crucial for LLM acceleration but remains a less-explored area.

To tackle the challenges mentioned above and facilitate the deployment of LLMs on edge devices, we offer the following contributions:

\begin{itemize}
\item{We propose \textbf{AccLLM} (as shown in Fig. \ref{fig:Overview}), a comprehensive LLM acceleration framework that leverages algorithm-hardware co-design to enable efficient and fast LLM inference on FPGAs.}
\item At the algorithm level, we introduce an aggressive compression algorithm, which integrates (1) pruning, (2) an innovative W$2$A$8$KV$4$ quantization that quantizes LLMs' weights to $2$-bit, activations to $8$-bit, and KV cache to $4$-bit, and (3) $\Lambda$-shaped attention, thus effectively enhancing computational efficiency, reducing memory and bandwidth requirements, while enabling long-sequence generation capabilities for LLMs.

\item{At the hardware level, we develop a dedicated FPGA accelerator featuring a reconfigurable computing engine to translate our algorithmic innovations into real hardware efficiency. Specifically, our accelerator is designed to accommodate: (1) both dense and sparse operations resulting from pruning, (2) diverse bit-widths ($2/4/8$-bit) introduced by our W2A8KV4 quantization, and (3) MM multiplications in the prefill stage and VM multiplications in the decode stage inherent in LLMs, as well as (4) the $\Lambda$-shaped attention integrated into our compression algorithm.}

\item{We conduct extensive experiments and ablation studies to validate the effectiveness of our AccLLM framework. Particularly, compared to the SOTA work FlightLLM \cite{Zeng2024FlightLLMEL} on Xilinx Alveo U280 FPGA, we achieve an \lyb{$\uparrow$$\mathbf{4.07}\times$} \lyb{energy efficiency} with \lyb{$\uparrow$$\mathbf{2.98}\times$} \lyb{throughput}.}
\end{itemize}

\lyb{The remainder of this paper is organized as follows: Sec. \ref{sec:related_work} reviews related works and Sec. \ref{sec:background_and_motivation} introduces preliminaries; Then, Sec. \ref{sec:algorithm} and Sec. \ref{sec:hardware} elaborate the algorithm and dedicated accelerator in AccLLM, respectively; Furthermore, Sec. \ref{sec:experiments} present extensive experiments and ablation studies, consistently validating AccLLM’s effectiveness; Finally, Sec. \ref{sec:conclusion} summarizes this paper.}




\section{Related Works}
\label{sec:related_work}

\subsection{Pruning for LLMs}
{Pruning boosts model compactness at the architectural level by removing redundant parameters, ranging from individual weights (unstructured pruning\cite{Shao2023OneShotSM, Syed2023PruneAT, Xu2024BESAPL}) to 
entire channels or layers (structured pruning\cite{Ma2023LLMPrunerOT, Liu2023DejaVC, Li2023LoSparseSC}).}
{Although unstructured pruning can achieve significant compression ratios, the resulting irregular sparsity is not conducive to hardware implementation \cite{Molchanov2016PruningCN}.}
In contrast, structured pruning is more compatible with hardware acceleration but often results in model accuracy degradation and limited sparsity \cite{He2017ChannelPF}. {To balance model accuracy and hardware efficiency, $N$:$M$ semi-structured pruning \cite{Frantar2023SparseGPTML, Sun2023ASA}, where $N$ out of every $M$ elements are pruned, is commonly adopted in prevalent LLMs \cite{Mishra2021AcceleratingSD, Zeng2024FlightLLMEL}.}
{For example, SparseGPT \cite{Frantar2023SparseGPTML} effectively prunes GPT-family models \cite{Zhang2022OPTOP} to achieve $2$:$4$ and $4$:$8$ sparsity in a one-shot manner without any retraining.}
Moreover, Wanda \cite{Sun2023ASA} leverages the product of weights and input activations to achieve $2$:$4$ semi-structured pruning, demonstrating improved perplexity in certain cases, such as Llama-2-13B \cite{Touvron2023Llama2O}.

\subsection{Quantization for LLMs}
{Quantization is a pivotal compression technique that converts floating-point values into discrete integers, thus enhancing LLMs' efficiency at the operator level.
It is typically categorized into two approaches: quantization-aware training (QAT) and post-training quantization (PTQ).
QAT \cite{Min2021MetaICLLT, Wei2021FinetunedLM, Liu2022FewShotPF} typically achieves higher quantization accuracy by fine-tuning the entire model using full training data, leading to substantial computational costs.
In contrast, PTQ \cite{Xiao2022SmoothQuantAA, Frantar2022GPTQAP} relies on only a small dataset for calibration, making it a more feasible solution for LLM quantization.}
{For example, GPTQ \cite{Frantar2022GPTQAP} introduces a one-shot weight quantization method using approximate second-order information, enabling the fast quantization of weights within GPT/OPT models to $3$ or $4$-bit with negligible accuracy degradation. To facilitate both weight and activation quantization, SmoothQuant \cite{Xiao2022SmoothQuantAA} employs equivalent mathematical transformations to shift the quantization complexity from activations to weights, allowing both to be quantized to $8$-bit. More recently, QServe \cite{Lin2024QServeWQ} integrates innovative progressive group quantization, smooth attention, and activation-aware channel reordering to achieve more aggressive quantization in a W$4$A$8$KV$4$ configuration ($4$-bit weights, $8$-bit activations, and $4$-bit KV cache).}

While these approaches demonstrate promising results in compressing LLMs while preserving performance, the residual computational and memory demands remain impractical for deployment on resource-constrained edge devices. This highlights the need for more aggressive model compression methods that combine orthogonal techniques, such as quantization and pruning, to produce even more compact LLMs.

\subsection{LLM Accelerators}
The remarkable performance of LLMs has driven efforts \cite{Zeng2024FlightLLMEL, Huang2024EdgeLLMAH, Xu2024LlamaFAE, Hong2022DFXAL} to deploy them in edge scenarios. 
One approach to achieve this goal involves integrating multiple edge devices into a unified system to enhance computational capacity and enable fast LLM acceleration. For instance, DFX \cite{Hong2022DFXAL} combines multiple FPGAs into a single large-scale accelerator, enabling low-latency, high-throughput inference for GPT-2 \cite{Radford2019LanguageMA}.
Another approach is to compress LLMs first and then design specialized accelerators tailored for the compact models. For example, LlamaF \cite{Xu2024LlamaFAE} uses quantization to compress both activations and weights of TinyLlama \cite{Zhang2024TinyLlamaAO} into $8$-bit formats and accelerates the resulting quantized MV multiplications with a fully pipelined accelerator. Moreover, FlightLLM \cite{Zeng2024FlightLLMEL} integrates quantization and pruning to compress LLMs and develops a dedicated accelerator with two key innovations: (1) a configurable sparse DSP chain optimized for diverse sparsity patterns to enhance computational efficiency, and (2) an always-on-chip decode scheme that reduces memory bandwidth requirements through low-precision quantization.

Despite the effectiveness of these accelerators, the substantial bandwidth demands of LLMs persistently limit achievable throughput, leading to under-utilized computing resources and underscoring the need for more aggressive compression algorithm. Additionally, these accelerators often lack adequate hardware support to address the scalability challenges associated with LLMs' long-sequence processing, impeding their deployment in real-world applications.

\section{Challenges and Motivations}
\label{sec:background_and_motivation}

\begin{figure}[!t]
\centering
\setlength{\abovecaptionskip}{0.1cm}
\includegraphics[width=0.99\columnwidth]{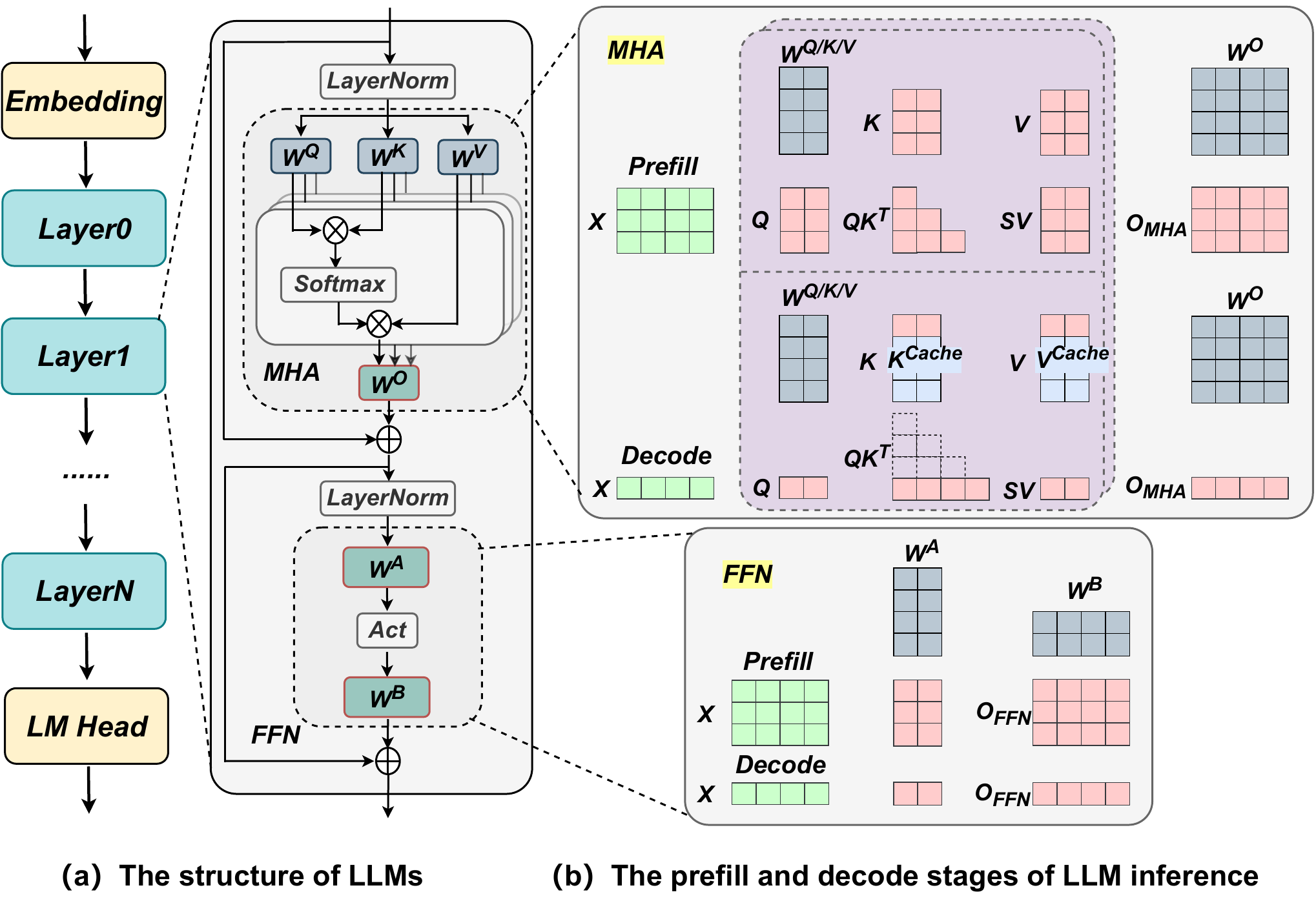}
\caption{(a) The structure of LLMs. (b) Illustrating the key computations during the prefill and decode stages of LLM inference.}
\label{fig:LLMArch}\vspace{-1em}
\end{figure}

In this section, we first outline the structure of LLMs and then explore three key challenges in LLM inference that motivate our proposed compression and acceleration framework.

\subsection{Structure of LLMs} 
\label{sec:LLM_structure}
Fig. \ref{fig:LLMArch} (a) illustrates the architecture of LLMs, which consists of a sequence of Transformer decoder layers, each comprising a multi-head attention ({MHA}) module and a feed-forward network ({FFN}) module. 
In the {\textbf{prefill stage}}, the input prompt is first embedded into $X$$\in$$\mathbb{R}^{l\times d}$, where $l$ represents the number of tokens and $d$ denotes the embedding dimension. This embedded matrix then serves as the input to Transformer layers. As depicted in Fig. \ref{fig:LLMArch} (b) (top), within each layer, the \textit{\textbf{MHA}} projects the input $X$ into the query ($Q_i$), key ($K_i$), and value ($V_i$) matrices for the $i^\text{{th}}$ attention head. This is achieved through three linear transformations using their respective weight matrices $W^Q_i$, $W^K_i$, and $W^V_i$:
\begin{equation}\label{eq:qkv}
Q_i=X\cdot W^Q_i, \ K_i=X\cdot W^K_i, \ V_i=X\cdot W^V_i. 
\end{equation}
Next, matrix $Q_i$ is first multiplied with $K_i^{\top}$ to compute the attention score $S_i$, which is then multiplied with $V_i$ to produce the attention output $A_i$ for the $i^\text{{th}}$ head. This process is expressed as follows, where $d_k$ denotes the hidden dimension:
 \begin{equation}\label{eq:QK}
S_i=\operatorname{Softmax}(S_i^{\prime})=\operatorname{Softmax}(\frac{Q_i K_i^{\top}}{\sqrt{d_k}}), \ A_i=S_iV_i.
\end{equation}
Finally, the attention output from all $h$ heads are concatenated and processed through the output linear projection layer $W^O$ to produce the final result of the {MHA} module $O_{\operatorname{MHA}}$:
\begin{equation}\label{eq:O}
    O_{\operatorname{MHA}}=\operatorname{concat}(A_1,..., A_d)\cdot W^O.
\end{equation}

As shown in Fig. \ref{fig:LLMArch} (b) (down), in the \textit{\textbf{FFN}} module, which consists of two linear layers ($W^{A}$ and $W^{B}$) separated by a non-linear activation function ($\sigma(\cdot)$), the computation for a given input $X$$\in$$\mathbb{R}^{l\times d}$ can be expressed as: 
\begin{equation}\label{eq:FFN}
     O_{\operatorname{FFN}}=\sigma(X\cdot W^{A}) \cdot W^{B}.
\end{equation}
where $O_{\operatorname{FFN}}$ is output of the {FFN} module. 

In summary, since the prefill stage processes the input prompt in parallel, the computations involved in both MHA (input and output linear projections in Eq. (\ref{eq:qkv}) and Eq. (\ref{eq:O}), respectively, and attention computations in Eq. (\ref{eq:QK})) as well as FFN (linear projections in Eq. (\ref{eq:FFN})) are \underline{\textit{MM}} multiplications.


In contrast, during the {\textbf{decode stage}}, the use of the KV cache eliminates repeated computations, allowing only one input token to be processed at a time. As a result, the computations in both MHA and FFN (Eqs. (\ref{eq:qkv})–(\ref{eq:FFN})) are reduced to \underline{\textit{VM}} multiplications.

\subsection{Challenges and Motivations}
\label{sec:observation}

\begin{table}[!t]
    \centering
    \setlength{\tabcolsep}{4pt}
    \caption{The Computational Complexity of Linear Operations\\ Across Different Stages in LLMs} \vspace{-0.6em}
    \renewcommand{\arraystretch}{1.2}
    \resizebox{\linewidth}{!}{
    \begin{threeparttable}{
    \begin{tabular}{c|ccc} \Xhline{3\arrayrulewidth}
         \textbf{Linear Operation} & \textbf{Formula}  & \textbf{Prefill Stage} & \textbf{Decode Stage} \\ \hline \hline
         \textbf{$\mathbf{Q}$/$\mathbf{K}$/$\mathbf{V}$/$\mathbf{O_{\operatorname{MHA}}}$ }  & $X W^Q, X W^K, X W^V, A W^O$  &  $4ld^2$ & $4d^2$  \\
         \textbf{Attention}  & $Q K^{\top}, SV$ & $l(l+1)d$ & $2(l+1)d$ \\ 
         \textbf{FFN}  & $\sigma(X W^{A})  W^{B}$ &  $2ldd_{\operatorname{FFN}}$  &  $2dd_{\operatorname{FFN}}$ \\ 
         \Xhline{3\arrayrulewidth}
    \end{tabular}}
    \begin{tablenotes}
		\footnotesize
		\item[] \textit{Notes:} $l$ represents the input sequence length, $d$ denotes the input feature dimension, and $d_{\text{FFN}}$ is the FFN hidden dimension.
	  \end{tablenotes}
    \end{threeparttable}}
    \label{tab:complexity}

\end{table}

\begin{figure}[!t]
\centering
\setlength{\abovecaptionskip}{0.1cm}
\includegraphics[width=\columnwidth]{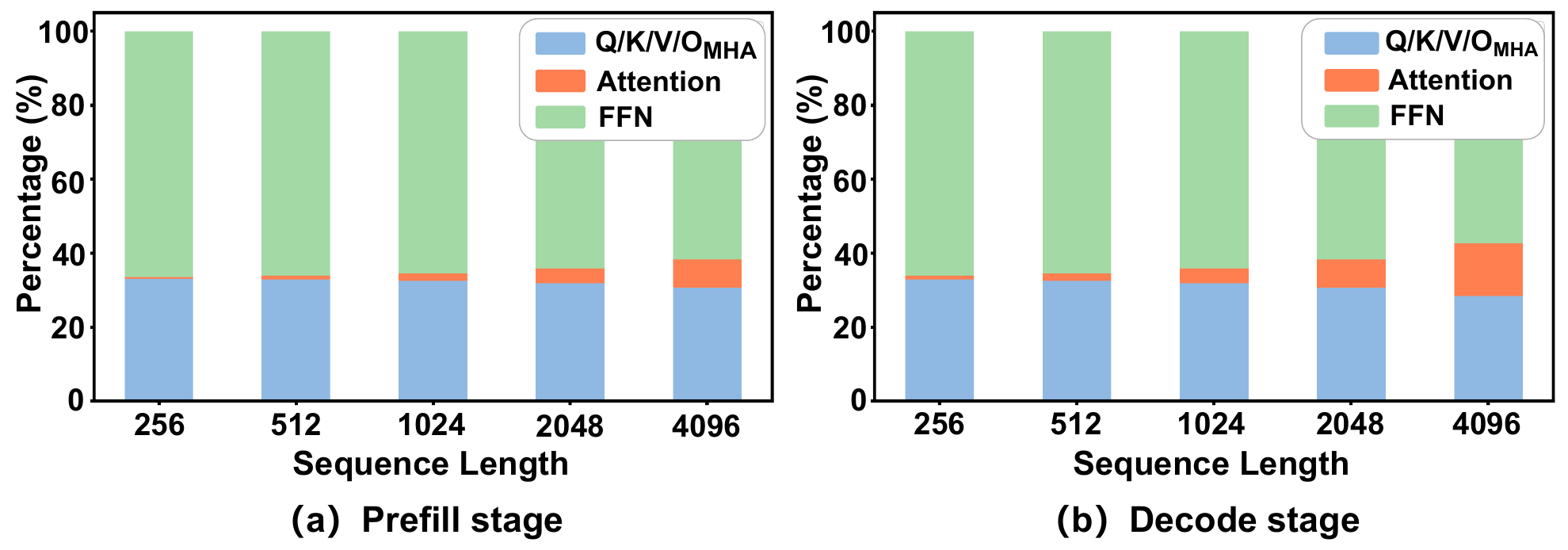}
\caption{The computational breakdown of Llama-2-7B \cite{Touvron2023Llama2O} inference during (a) the prefill stage and (b) the decode stage across different sequence lengths.}
\label{fig:historgram}\vspace{-1em}
\end{figure}

\subsubsection{Dominant Linear Layers}
\label{sec:motivation_1}
As outlined in Sec. \ref{sec:LLM_structure}, LLMs primarily consist of three types of operations: $Q$/$K$/$V$/$O_{\operatorname{MHA}}$ projections within MHAs defined in Eq. (\ref{eq:qkv}) and Eq. (\ref{eq:O}), attention computations in MHAs formulated in Eq. (\ref{eq:QK}), and linear projections within FFNs described in Eq. (\ref{eq:FFN}).
Their computational complexities are summarized in Table \ref{tab:complexity}. 
Using Llama-2-7B \cite{Touvron2023Llama2O} as an example, we analyze the computational breakdown during the prefill and decode stages for sequence lengths ranging from $256$ to $4096$.
As shown in Fig. \ref{fig:historgram}, the $\mathbf{Q}$/$\mathbf{K}$/$\mathbf{V}$/$\mathbf{O_{\operatorname{MHA}}}$ and \textbf{FFN linear layers} collectively account for over $\mathbf{90}\%$ of the computation. This dominance persists across varying sequence lengths and processing stages, underscoring the critical need to optimize linear layers for efficient deployment of LLMs on resource-constrained edge devices.

\begin{figure}[!t]
\centering
\setlength{\abovecaptionskip}{0.1cm}
\includegraphics[width=\columnwidth]{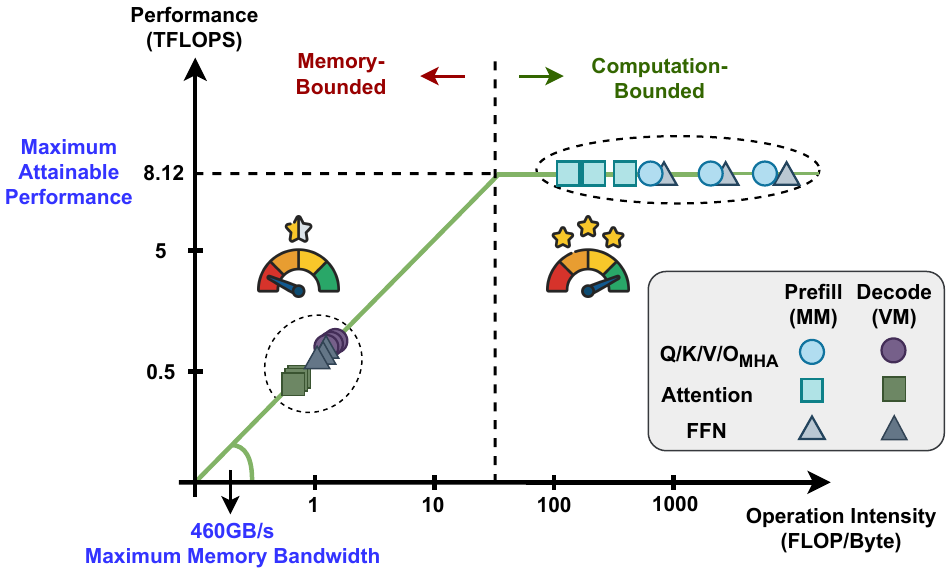}
\caption{Roofline analysis on the Xilinx Alveo U280 FPGA for the three primary types of linear operations during Llama-2-7B \cite{Touvron2023Llama2O} inference.}
\label{fig:roofline}\vspace{-1em}
\end{figure}

\subsubsection{Substantial Memory and Data Access Overhead Due to KV Cache in Attention Computations}
\label{sec:motivation_2}
As depicted in Fig. \ref{fig:historgram}, while attention map computations account for only a small fraction ($< 10\%$) of the total computations, they rely on the KV cache mechanism to store the $K$ and $V$ states of previous tokens, resulting in significant memory overhead, especially when handling long sequences.
For example, the memory requirements of KV cache for a $7$k-token context with Llama-2-7B could reach up to $3.5$GB\footnote{Memory requirements of KV cache = 2 * num of layers * sequence length * num of heads * head dimensions * bit-width of FP16 = $2*32*7k*32*128*16$b = $3.5$ GB}, which causes severe burdens for on-chip data buffering as well as off-chip data accessing, highlighting the need for optimization in attention computations.

\subsubsection{Memory-Bounded Decode Stage}
\label{sec:motivation_3}
In a single request, the prefill stage is executed only once, whereas the decoding stage is repeated for every output token in the response. Consequently, the decode stage often dominates the overall inference time, primarily due to the repeated loading of massive model parameters \cite{Liu2023DejaVC, Tang2024QuestQS}.
To further investigate the computation characteristics during LLMs' inference, we conduct a roofline analysis of Llama-2-7B \cite{Touvron2023Llama2O} on the Xilinx Alveo U280 FPGA platform.
As illustrated in Fig. \ref{fig:roofline}, we evaluate $\mathbf{Q}$/$\mathbf{K}$/$\mathbf{V}$/$\mathbf{O_{\operatorname{MHA}}}$, \textbf{Attention}, and \textbf{FFN} linear operations across the prefill and decode stages for three typical sequence lengths ($512$, $1024$, $2048$), where
the {operation intensity} $\mathcal{I}$
\cite{Kao2021FLATAO} is defined as the ratio of arithmetic operations to memory accesses:
\begin{equation}\label{eq:operation_intensity}
\mathcal{I}=\frac{\text { \# of Operations }}{\text { \# of Memory Accesses }}.
\end{equation}
High $\mathcal{I}$ indicates greater opportunities for data reuse, making performance limited by computational resources rather than bandwidth. Conversely, low $\mathcal{I}$ implies limited data reuse, leading to high bandwidth requirements and memory-bound performance \cite{Williams2008RooflineAI}.
As shown in Fig. \ref{fig:roofline}, in the prefill stage, where multiple input tokens are processed in parallel, the resulting MM multiplications facilitate data reuse and achieve peak performance. In contrast, during the decode stage, the autoregressive generation process introduces intensive VM multiplications with limited data reuse. This leads to under-utilization of computational resources and memory-bound performance, leading to an approximate $90\%$ drop in performance compared to the peak. 
In summary, the latency of LLM inference is predominantly determined by the decode stage, where performance is primarily limited by bandwidth. Thus, minimizing bandwidth requirements during the decode stage is essential for accelerating LLM inference.




\section{Aggressive Compression Algorithm}
\label{sec:algorithm}



In this section, we introduce an aggressive compression algorithm that is developed to effectively minimize computational and memory overhead while enhancing the long-sequence generation performance of LLMs.

\begin{figure*}[!t]
\centering
\setlength{\abovecaptionskip}{0.1cm}
\includegraphics[scale=.42]{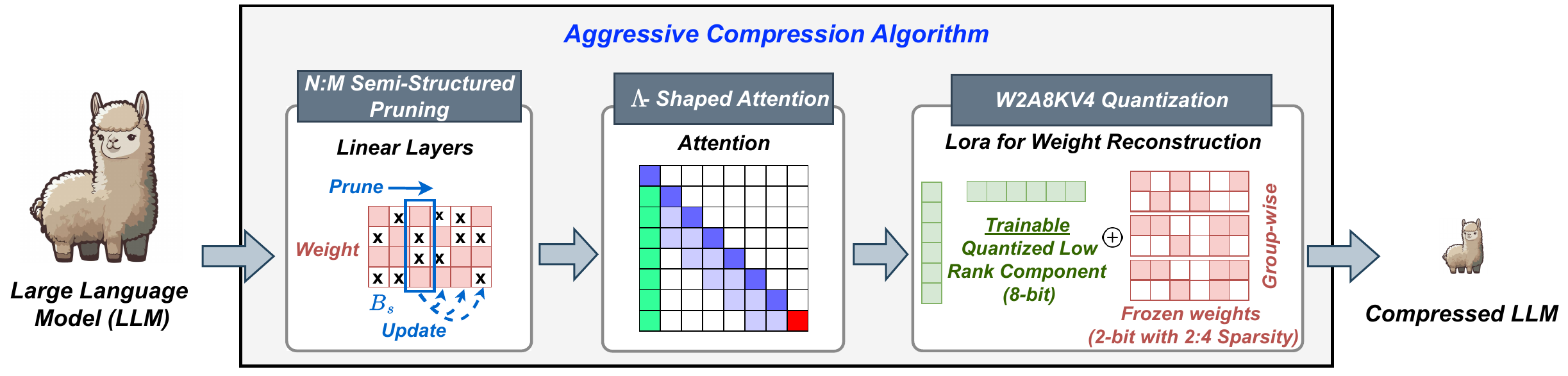}
\caption{The overview of our proposed aggressive compression algorithm, which includes (1) $N$:$M$ semi-structured pruning to reduce the {computational} complexity of {linear} layers, (2) $\Lambda$-shaped attention to facilitate both KV cache and LLMs' long-sequence generation, and (3) an innovative W2A8KV4 quantization technique to boost overall {throughput}.}
\label{fig:algflow}\vspace{-1.5em}
\end{figure*}

    

\begin{table}[!t]
    \centering
    \setlength{\tabcolsep}{0.8pt}
    \caption{Perplexity ($\downarrow$) of Llama-2-7B\cite{Touvron2023Llama2O} on the WikiText-103 dataset \cite{Merity2016PointerSM} with varying sequence lengths} \vspace{-1em}
    \renewcommand{\arraystretch}{1.2}
    \resizebox{\linewidth}{!}{
    \begin{threeparttable}{
    \begin{tabular}{c|ccc|c|ccccc} \Xhline{3\arrayrulewidth}
         \textbf{Model}   & \textbf{Pruning} & \textbf{$\Lambda$-Attention} & \textbf{Quantization} & {\textbf{\begin{tabular}[c]{@{}c@{}}Model  \\Size (GB) \end{tabular}}} & \textbf{3K}& \textbf{4K}& \textbf{5K}& \textbf{6K}& \textbf{7K}\\ \hline \hline
         \multirow{5}{*}{Llama-2-7B}  & \ding{55}   & \ding{55} & \ding{55} & 12.1 & 6.506  & 7.455 & 12.491  & 30.275 & 62.200    \\
           & \ding{51} & \ding{55}  & \ding{55} & 6.60 & 13.775 & 16.309 & 27.966 & 65.122 & 116.967 \\
           & \ding{55} & \ding{51}  & \ding{55} & 12.1  & 6.494  & 7.353 & 8.476 & 8.963 & 9.840 \\
           & \ding{55} & \ding{55}  & \ding{51} & 1.66 & 5.830 & 6.374 & 11.807 & 32.477 & 88.048 \\
           & \ding{51} & \ding{51}  & \ding{55} & 6.60 & 13.903 & 16.140 & 18.785 & 20.284 & 22.643 \\
           & \ding{51} & \ding{51}  & \ding{51} & 1.53 & 8.038  & 8.524 & 9.316 & 9.512 &  9.869
           \\\Xhline{3\arrayrulewidth}
    \end{tabular}}
    
    \end{threeparttable}}
    \vspace{-1.5em}
    \label{tab:perplexity}
\end{table}

    

\subsection{Overview}
\label{sec:alg_overview}

As illustrated in Fig. \ref{fig:algflow}, our aggressive compression algorithm combines three key techniques: (1) $2$:$4$ semi-structured pruning to reduce the \textit{computational} complexity of cost-dominant \textit{linear} layers, (2) $\Lambda$-shaped attention to simplify the \textit{attention} mechanism, thereby reducing the KV cache storage burden and improving the scalability of LLMs for long-sequence generation, and (3) an innovative W2A8KV4 quantization technique that boosts the \textit{throughput} of the memory-bounded decode stage while further reducing memory and data access overheads associated with the KV cache.

\subsubsection{$2$:$4$ Semi-Structured Pruning for Dominant Linear Layers}
\label{sec:method_pruning}
As discussed in \textbf{\textit{Sec. \ref{sec:motivation_1}}}, the linear layers dominate the computational workload across different sequence lengths and stages in LLM inference. To mitigate the substantial computational demands, we apply hardware-friendly $2$:$4$ semi-structured pruning \cite{Frantar2023SparseGPTML} to the weights of linear layers. 
As depicted in Fig. \ref{fig:algflow} (left), pruning is performed column-wise with a block size $B_s$. For $2$:$4$ semi-structured pruning, we set the block size $B_s=4$ and remove the $2$ least significant weights from each block based on the importance metric $S$:
\begin{equation}\label{eq:sparsegpt}
S_{i j}=\left[|W|^2 / \operatorname{diag}\left(H^{-1}\right)\right]_{i j},
H=X^{\top} X+\lambda I,
\end{equation}
where $i$ and $j$ represent the row and column indices of the weight matrix, respectively. $X$, $W$, and $\lambda$ denote the inputs, weights, and the Hessian dampening factor, respectively. 

After pruning each column, the weights in the remaining columns are updated to compensate for the pruning error {based on the optimal brain surgeon (OBS) approach \cite{Hassibi1993OptimalBS}}.

\subsubsection{$\Lambda$-Shaped Attention for KV Cache}
\label{sec:method_attetnion}
After resolving the dominant computational complexity of linear layers, we now focus on optimizing attention mechanisms. As outlined in \textbf{\textit{Sec. \ref{sec:motivation_2}}}, although attention computations constitute only a small portion of the total computations, the involved KV cache mechanism imposes substantial memory and data access overhead. To address this, we adopt $\Lambda$-shaped attention \cite{Xiao2023EfficientSL}, which is inspired by the ``attention sink'' phenomenon, highlighting that retaining the KV of initial tokens with strong attention scores can largely recover the performance of windowed attention. As shown in Fig. \ref{fig:algflow} (middle), by combining these initial tokens with high attention scores with the most recent tokens (window attention), this approach effectively reduces the storage burden of the KV cache while enhancing the scalability of LLMs for long-sequence generation.
For instance, when processing a $7$K-token context in Llama-2-7B, $\Lambda$-shaped attention reduces the memory requirements of the KV cache from $3.5$GB to $1$G, achieving a $\mathbf{71.4}\%$ reduction. Moreover, this method enables Llama-2-7B \cite{Touvron2023Llama2O} to handle longer sequences while maintaining promising perplexity results, as shown in Table \ref{tab:perplexity}. 


\begin{figure}[!t]
\centering
\setlength{\abovecaptionskip}{0.1cm}
\includegraphics[width=\columnwidth]{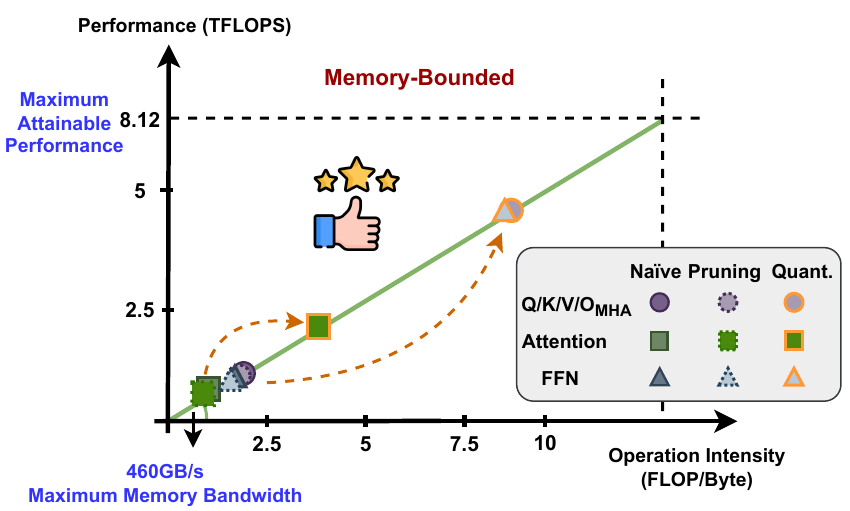}
\caption{Roofline analysis on the Xilinx Alveo U280 FPGA for the three primary types of linear operations during the decode stage of Llama-2-7B \cite{Touvron2023Llama2O} inference with different compression methods.}
\label{fig:roofline_prune_quant}\vspace{-1.5em}
\end{figure}

\subsubsection{W$2$A$8$KV$4$ Quantization for Memory-Bound Decode Stage}  
As discussed in \textbf{\textit{Sec. \ref{sec:motivation_1}}}, the latency of LLM
inference is predominantly constrained by bandwidth during the decode stage, due to the involved intensive memory-bounded VM multiplications. 
While the $2$:$4$ semi-structured pruning described in Sec. \ref{sec:method_pruning} effectively reduces \textit{computational} demands, it fails to alleviate bandwidth limitations or resolve the issue of low computation utilization, leading to limited \lyb{\textit{hardware performance (TFLOPS)}}. As demonstrated in Fig. \ref{fig:roofline_prune_quant}, where we perform a roofline analysis for the three primary types of linear operations during the decode stage of Llama-2-7B \cite{Touvron2023Llama2O} inference, \lyb{performance (TFLOPS)} remains largely unchanged even after pruning. To address this challenge, we provide an innovative {{W2A8KV4}} quantization method that compresses LLM weights to $2$-bit, activations to $8$-bit, and KV cache to $4$-bit. This approach offers two key benefits. \underline{\textit{First}}, as shown in Fig. \ref{fig:roofline_prune_quant}, it significantly enhance the throughput of the memory-bound decode stage by reducing the precision/bit-width of operations and thus increasing computational intensity. \underline{\textit{Second}}, it further minimizes memory and data access overheads associated with the KV cache. 
A detailed explanation of this approach is provided in the following section.

\begin{figure}[!t]
\centering
\setlength{\abovecaptionskip}{0.1cm}
\includegraphics[width=\columnwidth]{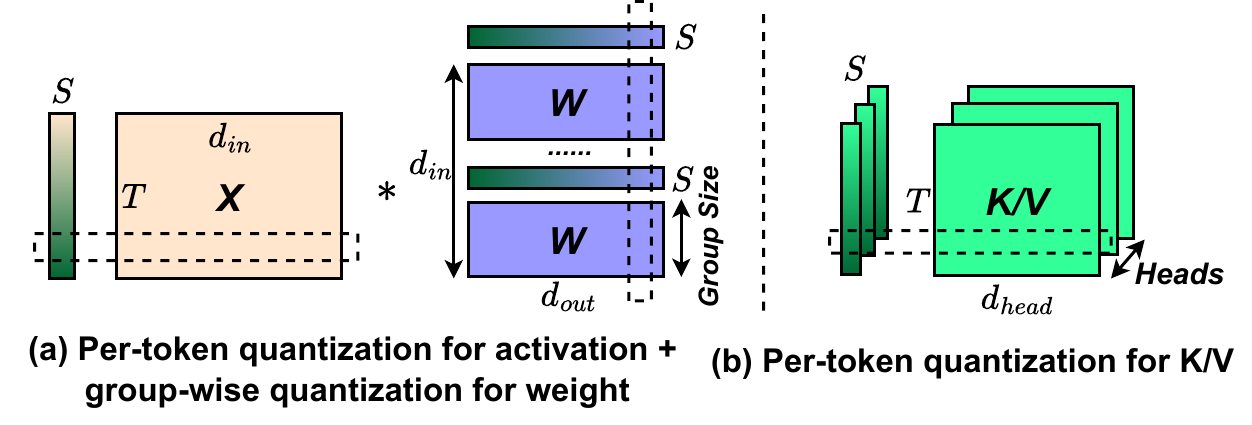}
\caption{The W2A8KV4 quantization: (a) per-token quantization for activation + group-wise quantization for weight and (b) per-token quantization for K/V.}
\label{fig:quantization}\vspace{-1.2em}
\end{figure}

\subsection{W$2$A$8$KV$4$ Quantization}
\label{sec:w2a8kv4}
\subsubsection{2-bit Weights}
\label{sec:alg_B_1}
To reduce the bandwidth requirements of dominant linear layers and enhance throughput during the memory-bounded decoding stage, we quantize the LLM weights to $2$-bit. 

{{{First}}}, to preserve performance in this low-bit configuration, we employ group-wise quantization \cite{Shao2023OmniQuantOC, Liao2024ApiQFO, lin2024awq}. As shown in Fig. \ref{fig:quantization} (a) (right), this method divides the weights $W^{i}$ within the $i^{\operatorname{th}}$ output channel into multiple groups, with all weights $W^{i,j}$ in the $j^{\operatorname{th}}$ group sharing a scaling factor $S^{i,j}$ and zero point $Z^{i,j}$:
\begin{equation}\label{eq:uniform_quantize}
W_Q^{i,j}=\operatorname{clip}(\lfloor W^{i,j} / S^{i,j} \rceil + Z^{i,j}, 0, 2^b-1), \ \text{where}
\end{equation}
\begin{small} 
\begin{equation}\label{eq:sf_zp}
S^{i,j}=\frac{W^{i,j}_{\operatorname{max}}-W^{i,j}_{\operatorname{min}}}{2^{b}-1}, \ Z^{i,j}=\operatorname{clip}\left(\left\lfloor-\frac{W^{i,j}_{\operatorname{min}}}{S^{i,j}}\right\rceil, 0,2^{b}-1\right). 
\end{equation}
\end{small}
Here, $W_Q$ represents the quantized weights, and $b$ is the quantization precision, set to $b=2$ in this case. To further enhance quantization performance, we integrate the learning weight clipping (LWC) method \cite{Liao2024ApiQFO, Shao2023OmniQuantOC, Liu2021NonuniformtoUniformQT} into our group-wise quantization approach. It introduces two learnable parameters, $\lambda$ and $\eta$, to facilitate quantization as follows:
\begin{equation}\label{eq:sf_zp_trainable}
X_{\operatorname{max}}=\sigma (\lambda)\operatorname{max}(X), \ X_{\operatorname{min}}=\sigma (\eta)\operatorname{min}(X),
\end{equation} 
where $\sigma$ represents the sigmoid function. 

\begin{table}[!t]
    \centering
    \setlength{\tabcolsep}{4pt}
    \caption{Perplexity ($\downarrow$) of Llama-2-7B \cite{Touvron2023Llama2O} on the WikiText-103 dataset \cite{Merity2016PointerSM} with a sequence length of 3K} \vspace{-1em}
    \renewcommand{\arraystretch}{1.2}
    \resizebox{0.8\linewidth}{!}{
    \begin{threeparttable}{
    \begin{tabular}{c|ccc|c} \Xhline{3\arrayrulewidth}
         \textbf{Model} & \textbf{Pruning}  & \textbf{W2} & \textbf{PEFT} & \textbf{WikiText-2} \\ \hline \hline
         \multirow{4}{*}{Llama-2-7B\cite{Touvron2023Llama2O}} & \ding{55}   & \ding{55} & \ding{55} &  6.506 \\
         & \ding{51}   & \ding{55} & \ding{55} & 13.775  \\
         & \ding{51}   & \ding{51} & \ding{55} & 16.695 \\
          & \ding{51} & \ding{51} & \ding{51} &  7.408  \\
          \Xhline{3\arrayrulewidth}
    \end{tabular}}
    
    \end{threeparttable}}
    \vspace{-1.5em}
    \label{tab:quantization}
\end{table}


Despite these efforts, achieving $2$-bit weight quantization in LLMs remains a significant challenge \cite{Frantar2022GPTQAP, lin2024awq}, and this difficulty is further exacerbated when combined with $2$:$4$ semi-structured pruning, as demonstrated in the third row of Table \ref{tab:quantization}. To mitigate the resulting performance degradation, inspired by the success of parameter-efficient fine-tuning (PEFT) \cite{Hu2021LoRALA, Dettmers2023QLoRAEF, Liao2024ApiQFO}, we further adopt LoRA fine-tuning to facilitate our pruning and quantization.

Specifically, we introduce a small set of learnable low-rank weights $A\in \mathbb{R}^{d_1 \times r}$ and $B\in \mathbb{R}^{d_2 \times r}$ ($r\ll d_2, d_1$) on top of the pruned and quantized weight matrix $\widetilde{W}_Q$ to approximate the original weights $W\in \mathbb{R}^{d_1 \times d_2}$:
\begin{equation}\label{eq:lora1}
W\approx \widetilde{W}_Q+A {B}^{T}.
\end{equation}
To stabilize fine-tuning, we initialize matrices $\widetilde{W}_Q$, $A$, and $B$ following \cite{Liao2024ApiQFO}:
\begin{equation}\label{eq:lora2}
 \underset{\lambda, \eta, A, B}{\operatorname{argmin}}\|\mathcal{F}(X, W)-\mathcal{F}(X, \widetilde{W}_Q,A,B)\|,
\end{equation}
where $X$ represents the input of block $\mathcal{F}$. This initialization ensures that the outputs of the pruned and quantized blocks closely align with those of the original blocks at the start of fine-tuning. During the subsequent fine-tuning process, $\widetilde{W}_Q$ is kept frozen in its $2$-bit representation to minimize fine-tuning cost, while $A$ and $B$ remain trainable to mitigate the performance degradation caused by pruning and quantization. 



It is worth noting that previous works \cite{Liao2024ApiQFO, Dettmers2023QLoRAEF} typically retain the low-rank component $A$ and $B$ in floating-point precision after fine-tuning. Although their size is relatively small, this approach necessitates the use of floating-point computing units. To eliminate this overhead, we quantize the well-trained $A$ and $B$ to $8$-bit before deployment, thus enabling their efficient processing in integers. 

Consequently, the original weights $W$ are approximated using $\widetilde{W}_Q$ and quantized low-rank components $A_Q$ and $B_Q$. The output of linear layers with input X can be expressed as:
\begin{equation}\label{eq:lora3}
XW \approx X(\widetilde{W}_Q+A_Q {B_Q}^{T})=X\widetilde{W}_Q+XA_Q {B_Q}^{T},
\end{equation}
This method has two primary advantages:
(1) As shown in Table \ref{tab:quantization}, this PEFT approach effectively mitigates the performance degradation caused by both quantization and pruning, successfully maintaining model performance even under aggressive compression settings;
(2) This promising performance comes at the cost of only a slight memory overhead. For example, in the case of Llama-2-7B, LoRA weights account for only $3\%$ of the original model’s weights.

\subsubsection{$8$-bit Activations and $4$-bit KV Cache}
To enable integer computations, we further employ $8$-bit per-token quantization to LLM activations. As depicted in Fig. \ref{fig:quantization} (a) (left), it assigns each activation token an individual scaling factor to enhance performance. 
Notably, to further reduce memory requirements of KV cache beyond the $\Lambda$-shaped attention introduced in Sec. \ref{sec:method_attetnion}, we apply $4$-bit per-token quantization for keys and values. Specifically, for each token, K/V elements of each head share a common scaling factor, as illustrated in Fig. \ref{fig:quantization} (b).


\section{FPGA-Based Reconfigurable Accelerator}
\label{sec:hardware}

In this section, we first introduce the overall hardware architecture in Sec. \ref{sec:hw_overall_arch}, followed by a detailed explanation of the reconfigurable computing engine in Sec. \ref{sec:hw_rce}. Finally, we illustrate the optimized dataflow in Sec. \ref{sec:hw_dataflow}, which facilitates both inter- and intra-layer pipelines.

\subsection{Micro-Architecture}

Our compressed LLMs involve four key computation types:
(1) MM multiplications during the prefill stage and MV multiplications in the decode stage;
(2) dense and sparse workloads from our $2$:$4$ semi-structured pruning;
(3) mixed-precision multiplications introduced by our W2A8KV4 quantization; and
(4) $\Lambda$-shaped attention for efficient KV cache.
To fully translate the algorithmic benefits into tangible hardware efficiency gains, we develop an FPGA-based reconfigurable accelerator, which incorporates:
(1) a Reconfigurable Computing Engine (RCE) to effectively support both MM and MV processing;
(2) a sparse selector to bypass zero elements and accelerate sparse workloads;
(3) a flexible DSP packing strategy for mixed-precision support; and
(4) an optimized dataflow to handle $\Lambda$-shaped attention.



\begin{figure}[!t]
\centering
\setlength{\abovecaptionskip}{0.1cm}
\includegraphics[width=\columnwidth]{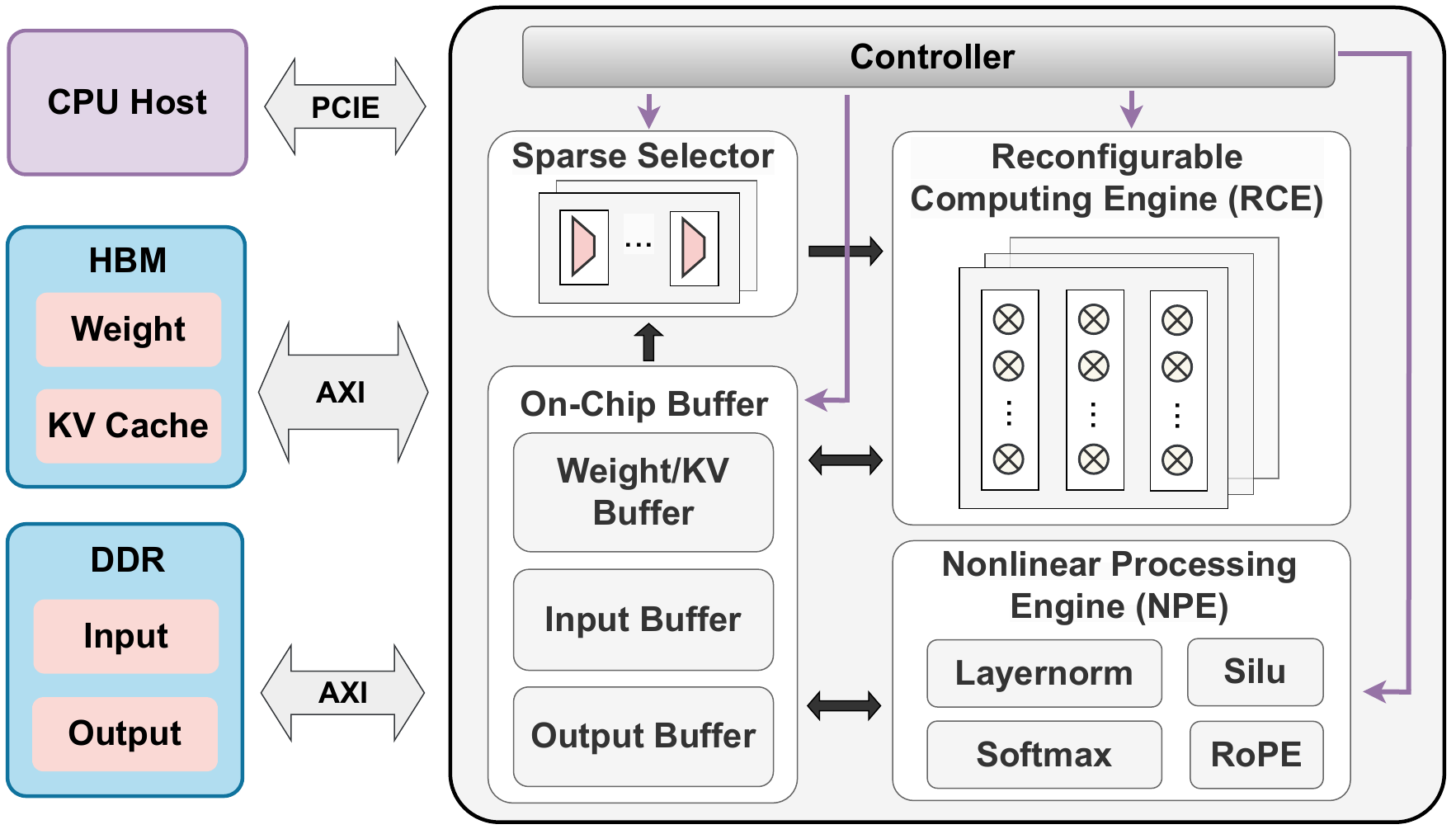}
\caption{The micro-architecture of our accelerator that integrates off-chip HBM/DDR interfaces, on-chip buffers, a sparse selector for sparsity support, a Reconfigurable Computing Engine (RCE) handling MM/VM multiplications, and a Nonlinear Processing Engine (NPE) for nonlinear operations.
}
\label{fig:hardware_arch}\vspace{-1em}
\end{figure}

\begin{figure*}[!t]
\centering
\setlength{\abovecaptionskip}{0.1cm}
\includegraphics[width=\textwidth]{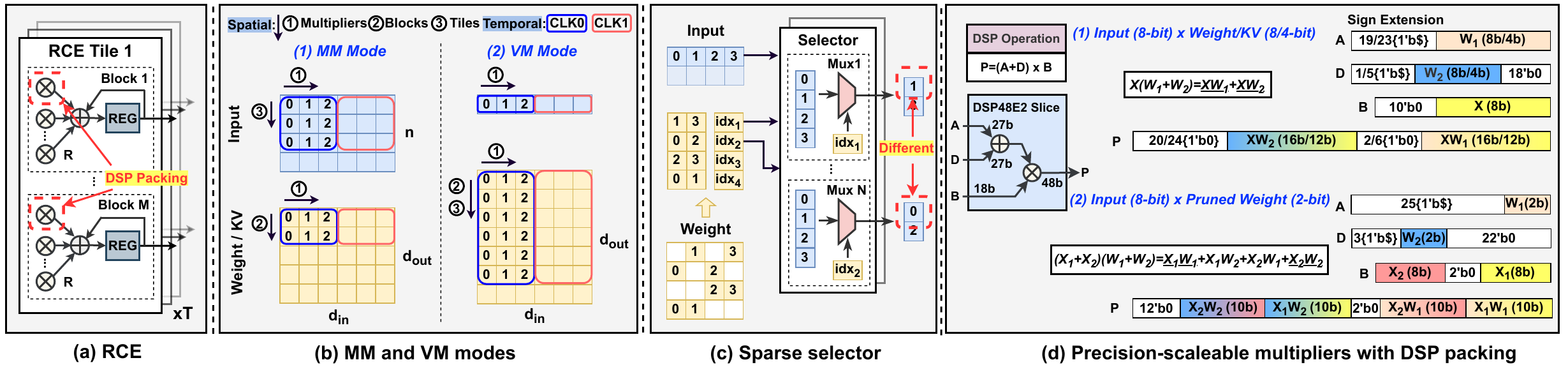}
\caption{(a) RCE architectural overview. (b) Reconfigurable operation modes (MM and MV modes) to support both MM and VM multiplications. (c) The incorporation of the sparse selector for $2$:$4$ sparsity. (d) The proposed flexible DSP packing strategy to support mixed-precision multiplications via precision-scalable multipliers.}
\label{fig:hardware_operation_mode}\vspace{-1.5em}
\end{figure*}


\label{sec:hw_overall_arch}

As shown in Fig. \ref{fig:hardware_arch}, our accelerator also includes: (1) a controller to manage global control signals, (2) on-chip buffer and external memory to store data (inputs, weights/KV values, and outputs), and (3) a Nonlinear Processing Engine (NPE) to execute nonlinear operations (e.g. softmax, activation function, and layernorm). Specifically, the external memory comprises High Bandwidth Memory (HBM) to store large single-access data, such as weights and KV cache, and DDR to store small single-access data, including input and output activations. 
Since nonlinear operations account for only a small part of total computations, the NPE processes them directly in floating-point to maintain accuracy. 

Next, we will detail our RCE and its integration with the sparse selector and flexible DSP packing
strategy to support the diverse computation workloads in compressed LLMs.

\subsection{Reconfigurable Computing Engine (RCE)}
\label{sec:hw_rce}
As shown in Fig. \ref{fig:hardware_operation_mode}, the RCE consists of $T$ processing tiles, each comprising $M$ PE blocks, and can be configured to compute both MM and VM multiplications. Furthermore, each PE block in RCE includes $R$ precision-scalable multipliers to support mixed-precision multiplications.
The RCE further incorporates a sparse selector composed of multiplexers to process weights with $2$:$4$ semi-structured pruning. 

\subsubsection{MM and VM Multiplications}
As shown in Fig. \ref{fig:hardware_operation_mode} (b), the RCE can be configured to operate in two distinct modes (MM and VM modes) for efficient execution of MM and VM multiplications, respectively.

\textbf{MM Mode:} To fully leverage the significant data reuse opportunities inherent in MM multiplication, we adopt an \textit{\textbf{input-output-parallel}} dataflow. As illustrated in Fig. \ref{fig:hardware_operation_mode} (b) (left), \underline{{\textit{multipliers}}} within each PE block perform computations along input channels, providing a parallelism of $R$ in input dimension. This enables partial sums to be directly accumulated across cycles within each block, thereby enhancing output reuse. Simultaneously, \underline{{\textit{blocks}}} within the same processing tile operate in parallel to handle different output channels of the weight, with input broadcast across blocks. This configuration achieves a parallelism of $M$ in output dimensions and improves input reuse. 
Additionally, \underline{{\textit{different processing tiles}}} process inputs from separate tokens simultaneously, with weights broadcast across tiles, facilitating weight reuse and achieving parallelism of $T$ in the token dimension.

\textbf{VM mode:} As the number of input tokens is reduced to 1, the weight reuse across different tokens in VM multiplication is no longer feasible. To maximize the available input and output reuse opportunity, we design a \textit{\textbf{output-parallel}} dataflow for VM. As illustrated in Fig. \ref{fig:hardware_operation_mode} (b) (right), \underline{{\textit{multipliers}}} within each PE block concurrently process weights in the same input channel, offering a parallelism of $R$ in input dimension. This also enables partial sums to be accumulated across cycles and thus enhances output reuse. \underline{{\textit{All PE blocks}}} within all processing tiles simultaneously process weights from different output channels, enabling parallelism of $M\times T$ in output dimension and facilitating input reuse. 

\subsubsection{Sparse and Dense Patterns}
To effectively handle sparse patterns and reduce redundant computations between inputs and pruned weights, inputs are first processed by the sparse selector before entering the RCE. 
As illustrated in Fig. \ref{fig:hardware_operation_mode} (c), the sparse selector identifies and selects relevant inputs based on the sparse indices of the pruned weights while discarding irrelevant ones, thus significantly enhancing computational efficiency. The sparse selector can be disabled when supporting dense patterns.

\subsubsection{Mixed-Precision Multiplications}
The W$2$A$8$KV$4$ quantization scheme described in Sec. \ref{sec:w2a8kv4} introduces three types of mixed-precision multiplications: (1) $8$-bit activations $\times$ $8$-bit LoRA weights, (2) $8$-bit query $Q$ $\times$ $4$-bit key $K$ and $8$-bit attention score $S$ $\times$ $4$-bit key $V$ in Eq. (\ref{eq:QK}), and (3) $8$-bit activations $\times$ $2$-bit weights. Meanwhile, as illustrated in Fig. \ref{fig:hardware_operation_mode} (b), the PE blocks within the same tile always operate in parallel to handle different output channels, allowing for effective input data reuse across blocks.
To efficiently execute these mixed-precision computations while fully leveraging this input reuse, we propose an \textit{\textbf{{flexible} DSP packing strategy}}. As shown in Fig. \ref{fig:hardware_operation_mode} (a), this strategy integrates two precision-scalable multipliers from adjacent blocks within the same processing tile into one DSP.

Specifically, as shown in Fig. \ref{fig:hardware_operation_mode} (d) (left), each DSP slice (DSP48E2) features a build-in $27\times 18$-bit multiplier, which performs the computation of $(A+D)\times B$, where $A$ and $D$ are $27$-bit values, and $B$ is a $18$-bit value. 
To effectively handle (1) {$8$-bit activation $\times$ $8$-bit LoRA weights},
as illustrated in Fig. \ref{fig:hardware_operation_mode} (d-1), we map two weights $W_{1}$ and $W_{2}$ to $A$ and $D$, respectively, and activations $X$ to $B$. 
This enables a single DSP to efficiently compute two multiplications via $X(W_{1}+W_{2})=XW_{1}+XW_{2}$, thus greatly saving DSP consumption.
Similarly, regarding (2)$8$-bit $Q/S$ $\times$ $4$-bit $K/V$, we treat the $4$-bit $K/V$ as $W$ and the $8$-bit $Q/S$ as $X$, thus allowing for the packing of two multiplications using identical DSP data path routing.

For (3) {$8$-bit activations $\times$ $2$-bit weights}, additional optimizations are required. Specifically,
since $2$-bit quantization for linear layer weights is always paired with pruning, the inputs are first processed by the sparse selector, which selects relevant inputs based on the pruned weights. However, as the pruned inputs typically vary between different weight rows (as indicated by the red dotted line in Fig. \ref{fig:hardware_operation_mode} (c)), the input reuse opportunities between adjacent PE blocks in RCE are eliminated. 
To overcome this limitation, we pack two inputs $X_1$ and $X_2$ into $B$, while mapping two weights $W_{1}$ and $W_{2}$ to $A$ and $D$, respectively, as shown in Fig. \ref{fig:hardware_operation_mode} (d-2). This enables a single DSP to execute four multiplications by $(X_1+X_2)(W_{1}+W_{2})=X_{1}W_{1}+X_{1}W_{2}+X_{2}W_{1}+X_{2}W_{2}$.
The required results (i.e., $X_1W_1$ and $X_2W_2$) are then selectively extracted as the final output. As a result, the proposed flexible DSP packing strategy significantly enhances DSP utilization efficiency.

\subsection{Dataflow Optimization}
\label{sec:hw_dataflow}

\begin{figure}[!t]
\centering
\setlength{\abovecaptionskip}{0.1cm}
\includegraphics[width=1\columnwidth]{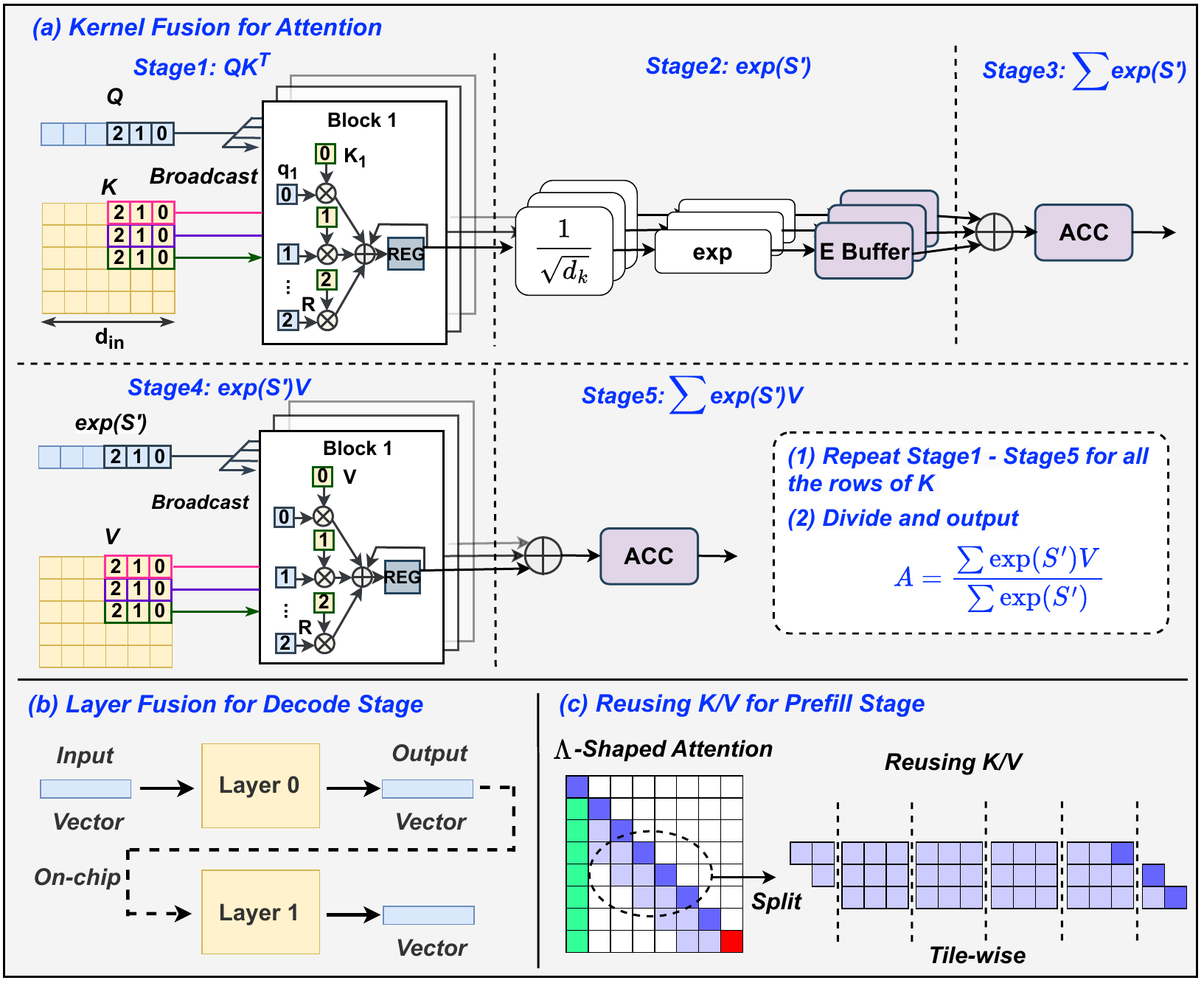}
\caption{Dataflow optimization with (a) kernel fusion for attention (in both prefill and decode stage), (b) layer fusion for decode stage, and (c) reusing K/V for prefill stage.}
\label{fig:kernelfusion}\vspace{-1.7em}
\end{figure}

\subsubsection{Kernel Fusion for Attention} The standard implementation of {{attention}} in Eq. (\ref{eq:QK}) involves a \textit{sequential} three-step computations - $Q K^T$, $\operatorname{Softmax}(\cdot)$, and $SV$ - primarily due to the row-wise data dependency of the $\operatorname{Softmax}$ operation, which requires accumulating $H$ datas in one row before performing the element-wise division:

\begin{equation}\label{eq:softmax}
\operatorname{Softmax}(x_i)=\frac{\operatorname{exp}(x_i)}{\sum_{l=0}^H \operatorname{exp}(x_l)},
\end{equation}
where $x_i$ represents an element within the row.

Since on-chip buffer is typically insufficient for storing all intermediate results in \textit{prefill} stage, it leads to redundant off-chip data accesses. To address this issue and enhance performance, inspired by \cite{Dao2022FlashAttentionFA, Bai2024SWATSA}, 
we fuse these computations into a single operation by reinterpreting $\operatorname{Softmax}$ operation:   

\begin{equation}\label{eq:kernelfusion}
\begin{aligned}
A_{j, k} & =\sum_{n=0}^H S_{j, n} V_{n, k}=\sum_{n=0}^H \frac{\operatorname{exp} \left(S_{j, n}^{\prime}\right)}{\sum_{l=0}^H \operatorname{exp} \left(S_{j, l}^{\prime}\right)} V_{n, k} \\
& =\left(\frac{1}{\sum_{l=0}^H \operatorname{exp} \left(S_{j, l}^{\prime}\right)}\right)\left(\sum_{n=0}^H \operatorname{exp} \left(S_{j, n}^{\prime}\right) V_{n, k}\right),
\end{aligned}
\end{equation}
where $A_{j, k}$ is the element at $j^{th}$ row and $k^{th}$ column of the final attention output.

As shown in Fig. \ref{fig:kernelfusion} (a), using a single row of $Q$ as an example, the computations in Eq. (\ref{eq:kernelfusion}) mainly consists of five stages. In {Stage $\mathbf{1}$}, $QK^T$ is computed for several rows of $K$. The results are subsequently processed by the NPE in {Stage $\mathbf{2}$ and $\mathbf{3}$} to obtain $ \operatorname{exp} (S^{\prime})$ and $\sum \operatorname{exp} (S^{\prime})$, respectively. Further, the $ \operatorname{exp} (S^{\prime})$ will be multiplied by $V$ in {Stage $\mathbf{4}$} and accumulated to obtain $\sum \operatorname{exp} (S^{\prime})V$ in {Stage $\mathbf{5}$}. These five stages are repeated until all rows of $K$ are processed, generating the final $\sum \operatorname{exp} (S^{\prime})V$ and $ \sum \operatorname{exp} (S^{\prime})$. 
Finally, the output is obtained by dividing $\sum \operatorname{exp} (S^{\prime})V$ by $\sum \operatorname{exp} (S^{\prime})$.

This rearrangement allows the multiplication of $V$ to be executed before completing the division, thus enabling computation fusion and reducing data access costs. 
While the on-chip buffer is sufficient to store intermediate results (vector) during the {decode} stage, this kernel fusion technique also facilitates pipelining processing within attention computations. Therefore, we apply kernel fusion in both the prefill and decode stages to improve performance.


\subsubsection{Layer Fusion for the Decode Stage}
In the decode stage, the input and output activations are small vectors rather than large matrices, allowing them to be entirely stored within the on-chip buffer of the FPGA. To minimize off-chip memory access, we fuse computations of all layers in this stage by directly using the output of the current layer as the input to the subsequent layer, as illustrated in Fig. \ref{fig:kernelfusion} (b).

\subsubsection{Reusing K/V for the Prefill Stage}
As illustrated in Fig. \ref{fig:kernelfusion} (c), the $\Lambda$-shaped attention patterns between adjacent rows exhibit a one-token shift overlap, offering an opportunity to reuse $K$ and $V$ data during the prefill stage. As such, we vertically split the attention into several tiles based on the size of RCE in our accelerator and process them sequentially. Within each tile, multiple attention rows are computed in parallel while maintaining shared access to the KV data.

\section{Experiments}
\label{sec:experiments}

\subsection{Experimental Setup} 

\subsubsection{Model, Datasets, Algorithm Setup, and Baselines}
\textbf{\textit{{Model and Datasets}}}: We evaluate our aggressive compression algorithm using the widely adopted LLM, Llama-2-7B \cite{Touvron2023Llama2O}, on the commonly used WikiText-103 and WikiText-2 \cite{Merity2016PointerSM} datasets and report their perplexity. 
\textbf{\textit{{Algorithm Setup}}}:
Our aggressive compression algorithm combines three key techniques: (1) $2$:$4$ semi-structured pruning to reduce the {computational} complexity of cost-dominant {linear} layers, (2) $\Lambda$-shaped attention to simplify the {attention} mechanism, and (3) an innovative W2A8KV4 quantization technique that reduces memory and data access overheads.
As for (1) {$2$:$4$ semi-structured pruning,} we follow \cite{Frantar2023SparseGPTML} to use $128$ randomly sampled 2048-token segments from the first shard of the C4 dataset \cite{Raffel2019ExploringTL} as the calibration data.
For {(2) $\Lambda$-shaped attention mechanism,} we set the KV cache size to 2048, consisting of 4 initial tokens and 2044 most recent tokens \cite{Xiao2023EfficientSL}. 
Regarding (3) W2A8KV4 quantization, we set the group size to $64$ for group-wise quantization for weight. For the quantization initialization process in Eq. (\ref{eq:lora2}), we follow \cite{Liao2024ApiQFO} and randomly sample $128$ sentences from the training set of WikiText-103 and WikiText-2 \cite{Merity2016PointerSM}, which serve as calibration datasets. Subsequently, we perform dataset-specific LoRA fine-tuning on their respective datasets. \textbf{\textit{{Baselines}}}:
We compare our compressed algorithm with four counterparts: (1) the half-precision (FP16) baseline, (2) the widely used LLM quantization work, SmoothQuant \cite{Xiao2022SmoothQuantAA}, (3) the SOTA W$4$A$8$KV$4$ LLM quantization framework, QoQ \cite{Lin2024QServeWQ}, and (4) the widely adopted LLM pruning method, SparseGPT \cite{Frantar2023SparseGPTML}, in terms of perplexity on varying sequence lengths and model size after compression.

\subsubsection{Accelerator Setup and Baselines}
\textbf{\textit{{Hardware Setup}}}:
The parallelism of the reconfigurable computing engine in our accelerator $(R\times M) \times T$ (see Fig. \ref{fig:hardware_operation_mode} (a)) is configured as $(32 \times 16) \times 16$.
Our dedicated accelerator, AccLLM, is coded with Verilog and synthesized with the Vivado Design Suite. We evaluate its performance on the Xilinx Alveo U280 FPGA at a clock frequency of $225$MHz.
Table \ref{tab:hw_resources} summarizes the U280 resource consumption of AccLLM. Additionally, we follow \cite{Dass2022ViTALiTyUL,Shi2024TrioViTPQ} to develop a cycle-accurate simulator for our accelerator to provide fast and reliable performance estimations, which are validated against the RTL implementation to ensure correctness.
\textbf{\textit{{Baselines}}}: We compare AccLLM with (1) half-precision (FP16) Llama-2-7B \cite{Touvron2023Llama2O} on NVIDIA A100 GPU, (2) mixed-quantized and sparse Llama-2-7B \cite{Touvron2023Llama2O} on the SOTA FPGA accelerator, FlightLLM \cite{Zeng2024FlightLLMEL}, and (3) W$4$A$16$ quantized and sparse ChatGLM2-6B \cite{Zeng2024ChatGLMAF} on its dedicated edge accelerator, EdgeLLM \cite{Huang2024EdgeLLMAH}. We compare with them in terms of \lyb{throughput, power, and energy efficiency}.

\begin{table}[]
\centering
\setlength{\tabcolsep}{2pt}
\caption{Resource consumption of our dedicated accelerator} \vspace{-1em}
\renewcommand{\arraystretch}{1.1}
\resizebox{0.85\linewidth}{!}{
\begin{tabular}{c|cccc}
\Xhline{3\arrayrulewidth}
\textbf{Resources} & \textbf{BRAM} & \textbf{DSP}  & \textbf{LUT}    & \textbf{FF}     \\ \hline \hline
\textbf{Available} & 2016  & 9024 & 1304K & 2607K\\
\rowcolor{dark-green!16} \textbf{Used}      & 513 (25.4\%)  & 4497 (49.8\%) & 420K(32.2\%) & 274K(10.5\%) \\ \Xhline{3\arrayrulewidth}
\end{tabular}} \label{tab:hw_resources} \vspace{-1.7em}
\end{table}


\begin{table}[!t]
    \centering
    \setlength{\tabcolsep}{0.5pt}
    \caption{Performance of Llama-2-7B \cite{Touvron2023Llama2O} on the WikiText-103  dataset \cite{Merity2016PointerSM} with varying sequence lengths under different compression algorithms} \vspace{-0.8em}
    \renewcommand{\arraystretch}{1.2}
    \resizebox{\linewidth}{!}{
    \begin{threeparttable}{
    \begin{tabular}{cccccccc} \Xhline{3\arrayrulewidth}
          \multirow{2}{*}{\textbf{Method}}  & \multirow{2}{*}{\textbf{Algorithm}} & \multirow{2}{*}{\textbf{\begin{tabular}[c]{@{}c@{}}Model Size \\ (GB) \end{tabular}}}& \multicolumn{5}{c}{\textbf{Perplexity ($\downarrow$)}} 
            \\ \cline{4-8}
          & & &\textbf{3k} &\textbf{4k} &\textbf{5k} &\textbf{6k} &\textbf{7k}
          \\ \hline \hline
           FP16 & - & 12.1 & 6.506 & 7.455 & 12.491 & 30.275 & 62.200 \\ \hline
           W8A8 & SmoothQuant & 6.03 & 6.778 & 7.743 & 13.090 & 32.478 & 66.430 \\ \hline
           \multirow{1}{*}{W4A8KV4} & QoQ & 3.08 & 7.142 & 8.186 & 13.707 & 33.729 & 67.240
          
             \\ \hline
          2:4 Pruning & SparseGPT & 6.60 & 13.775 & 16.309 & 27.966 & 65.122 & 116.967 \\ \hline \rowcolor{dark-green!16}
          \makecell{W2A8KV4 \\+ 2:4 Pruning\\+ $\Lambda$-Shaped Attention } & Ours & 1.53 & 8.038 & 8.524 & 9.316 & 9.512 & 9.869

          \\
          \Xhline{3\arrayrulewidth}
    \end{tabular}}
    
    \end{threeparttable}}
    \vspace{-1.5em}
    \label{tab:alg_perplexity_wikitext103}
\end{table}
 
          
           
    

\begin{table}[!t]
    \centering
    \setlength{\tabcolsep}{0.5pt}
    \caption{Performance of Llama-2-7B \cite{Touvron2023Llama2O} on the WikiText-2 dataset \cite{Merity2016PointerSM} with varying sequence lengths under different compression algorithms} \vspace{-0.8em}
    \renewcommand{\arraystretch}{1.2}
    \resizebox{\linewidth}{!}{
    \begin{threeparttable}{
    \begin{tabular}{cccccccc} \Xhline{3\arrayrulewidth}
 \multirow{2}{*}{\textbf{Method}}  & \multirow{2}{*}{\textbf{Algorithm}} & \multirow{2}{*}{\textbf{\begin{tabular}[c]{@{}c@{}}Model Size \\ (GB) \end{tabular}}}& \multicolumn{5}{c}{\textbf{Perplexity ($\downarrow$)}} 
            \\ \cline{4-8}
          & & &\textbf{3k} &\textbf{4k} &\textbf{5k} &\textbf{6k} &\textbf{7k}
          \\ \hline \hline
           FP16 & - & 12.1 &  18.497 & 20.608 & 30.619 & 63.461 & 114.484 \\ \hline
           W8A8 & SmoothQuant & 6.03 & 18.967 & 21.246  & 31.892  &  67.059 & 120.419  \\ \hline
           \multirow{1}{*}{W4A8KV4} & QoQ & 3.08 &  41.220 & 44.845  & 62.171  & 113.396 & 180.845
          
             \\ \hline
          2:4 Pruning & SparseGPT & 6.60 & 54.516 & 67.892 & 102.321 & 194.244 & 317.622 \\ \hline \rowcolor{dark-green!16}
          \makecell{W2A8KV4  \\+ 2:4 Pruning\\+ $\Lambda$-Shaped Attention} & Ours & 1.53 & 10.992 & 11.857 & 12.101 & 12.502 & 12.669
           
          \\
          \Xhline{3\arrayrulewidth}
    \end{tabular}}
    
    \end{threeparttable}}
    \vspace{-2em}
    \label{tab:alg_perplexity_wikitext2}
\end{table}

\begin{figure}[!t]
\centering
\setlength{\abovecaptionskip}{-0.1cm}
\includegraphics[width=1\columnwidth]{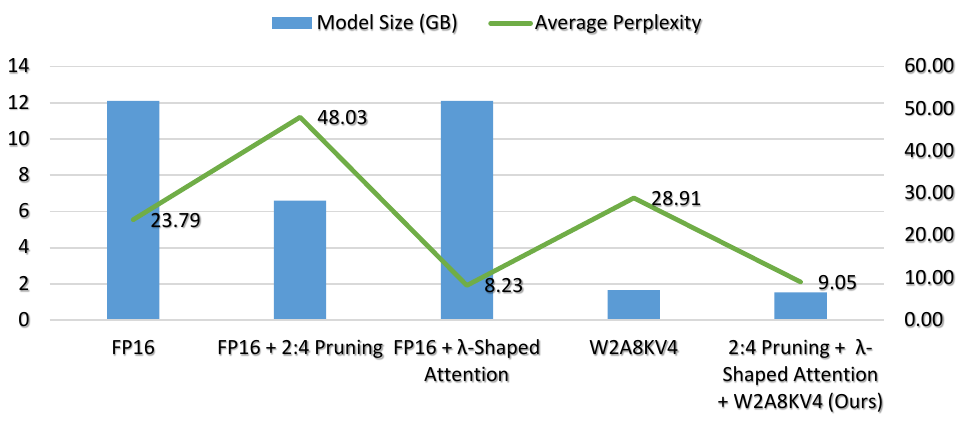}
\caption{The ablation studies on model size and average perplexity across 3k-7k lengths of Llama-2-7B \cite{Touvron2023Llama2O} on the WikiText-103 dataset \cite{Merity2016PointerSM}.}
\label{fig:ppl_histogram}\vspace{-1.7em}
\end{figure}

\subsection{Advancement of Our Aggressive Compression Algorithm}
Tables \ref{tab:alg_perplexity_wikitext103} and \ref{tab:alg_perplexity_wikitext2} show the performance of Llama-2-7B on WikiText-103 and WikiText-2 datasets \cite{Merity2016PointerSM} under different LLM compression methods. We can draw the following conclusions: 
\textbf{{(1)} Superior compression efficiency:} Our aggressive compression algorithm achieves the smallest compressed model size ($\mathbf{1.53}$ GB), which is only $\mathbf{12.6}\%$ of the FP16 baseline. When compared with SOTA quantization and pruning baselines, our approach can reduce the model size by \lyb{$\mathbf{50.3}\%$$\sim$$\mathbf{76.8}\%$}. This remarkable efficiency is attributed to the progressive combination of the innovative W$2$A$8$KV$4$ quantization and $2$:$4$ semi-structured pruning.
\textbf{{(2) Better compression-accuracy trade-offs on short sequences}}:
On the WikiText-103 dataset, our method achieves better trade-offs between compression ratio and model performance when compared with SmoothQuant \cite{Xiao2022SmoothQuantAA} and QoQ \cite{Lin2024QServeWQ}. Specifically, we reduce the model size by $\mathbf{74.6}\%$ and $\mathbf{50.3}\%$ compared to SmoothQuant and QoQ, while incurring only a negligible increase in perplexity of $\mathbf{1.26}$ and $\mathbf{0.896}$, respectively, under the $3$k sequence length. Furthermore, compared to the pruning baseline SparseGPT \cite{Frantar2023SparseGPTML}, our method not only achieves a $\mathbf{76.8}\%$ reduction in model size but also offers lower perplexity ($\downarrow$$\mathbf{5.737}$) under the $3$k sequence length.
On the WikiText-2 dataset, our approach consistently achieves the lowest model size and perplexity among all baselines, which further validates the effectiveness of our algorithm optimizations.
\textbf{{(3)} Exceptional performance on long sequences:} On the WikiText-103 dataset, while our method incurs a slight perplexity increase over FP16 and quantization baselines for short sequences ($\leq 4$k tokens), it surpasses all baselines on long sequences ($\geq 5$k tokens), highlighting the effectiveness of our adopted $\Lambda$-shaped attention technique. The benefits are even more pronounced on the WikiText-2 dataset, where our approach achieves a perplexity reduction of up to  $\mathbf{101.815}$$\sim$$\mathbf{304.953}$ under the $7$K token sequence, outperforming all baselines.


\textbf{Algorithm Ablation Studies:} We further validate the effectiveness of each component in our proposed aggressive compression algorithm by evaluating its impact on model size and average perplexity across sequence lengths ranging from $3$k-$7$k. As illustrated in Fig. \ref{fig:ppl_histogram}, we observe the following: {(1)} The $2$:$4$ semi-structured pruning eliminates redundant parameters, shrinking the model size by \lyb{$\mathbf{45.5}\%$} compared to the FP16 baseline, but this comes at the cost of a \lyb{$\mathbf{24.24}$} increase in average perplexity. 
{(2)} Although not directly reducing the model size, $\Lambda$-shaped attention significantly enhances the model’s ability to handle long sequences, leading to a \lyb{$\mathbf{15.56}$} decrease in average perplexity compared to the FP16 version. {(3)} The innovative W$2$A$8$KV$4$ quantization with LoRA fine-tuning achieves an impressive \lyb{$\mathbf{86.3}\%$} compression while maintaining comparable perplexity to the FP16 baseline. {(4)} The combination of these three techniques yields an \lyb{$\mathbf{87.4}\%$} overall reduction in model size while preserving performance. 

\begin{table}[!t]
\centering
\setlength{\tabcolsep}{1.pt}
\caption{Comparisons with SOTA transformer accelerators} \vspace{-0.9em}
\renewcommand{\arraystretch}{1.2}
\resizebox{\linewidth}{!}{
\begin{threeparttable}{
\begin{tabular}{c|c|c|c|c|c}
    \Xhline{3\arrayrulewidth}
    \textbf{Accelerator} &                        \textbf{\begin{tabular}[c]{@{}c@{}}GPU \end{tabular}}&   \textbf{EdgeLLM \cite{Huang2024EdgeLLMAH}}&   \multicolumn{2}{c|}{\textbf{FlightLLM \cite{Zeng2024FlightLLMEL}}}  & \textbf{\textcolor{dark-green} {Ours}}  \\ \hline \hline
    \textbf{Device}  & \textbf{\begin{tabular}[c]{@{}c@{}}NVIDIA \\ A100 GPU \end{tabular}} & \textbf{\begin{tabular}[c]{@{}c@{}}Xilinx \\ VCU128 \end{tabular}}& \textbf{\begin{tabular}[c]{@{}c@{}}Xilinx Versal\\ VHK158 \end{tabular}}& \textbf{\begin{tabular}[c]{@{}c@{}}Xilinx Alveo \\ U280 \end{tabular}} &  \textbf{\begin{tabular}[c]{@{}c@{}}Xilinx Alveo \\ U280 \end{tabular}} 

    \\ \hline
    \textbf{\begin{tabular}[c]{@{}c@{}}Frequency\\ (MHz)\end{tabular}} & \textbf{1410} & \textbf{125}& \textbf{225} & \textbf{225}  & \textbf{225}      

    \\ \hline
    \textbf{Model} & \textbf{Llama-2-7B \cite{Touvron2023Llama2O}}& \textbf{ChatGLM2-6B \cite{Zeng2024ChatGLMAF}}& \multicolumn{3}{c}{\textbf{Llama-2-7B \cite{Touvron2023Llama2O}}}  

    \\ \hline
    \textbf{DSP Used} & - & 5587 & - & 6345 & \cellcolor{dark-green!15} \textbf{{4497}} 


    \\
    \hline

    \textbf{\begin{tabular}[c]{@{}c@{}}Throughput\\ (Token/s)\end{tabular}} &  45 &  75 & 92.5 & 55 & \cellcolor{dark-green!15} \textbf{{164}}     \\ 
    \hline
    \textbf{Power (W)}& 220 & 50.77 & 155 & 45 & \cellcolor{dark-green!15} \textbf{{33}}    
    \\ 
    \hline
    \textbf{\begin{tabular}[c]{@{}c@{}}Energy Efficiency\\ (Token/J)\end{tabular}} & 0.2  & 1.47 & 0.6 & 1.22 & \cellcolor{dark-green!15} \textbf{4.96}

    \\  \Xhline{3\arrayrulewidth}                                                             
    \end{tabular}}  
    \end{threeparttable}}
    \vspace{-1.5em}
    \label{tab:hw_accelerator}
\end{table}

\begin{figure}[!t]
\centering
\setlength{\abovecaptionskip}{-0.1cm}
\includegraphics[width=1\columnwidth]{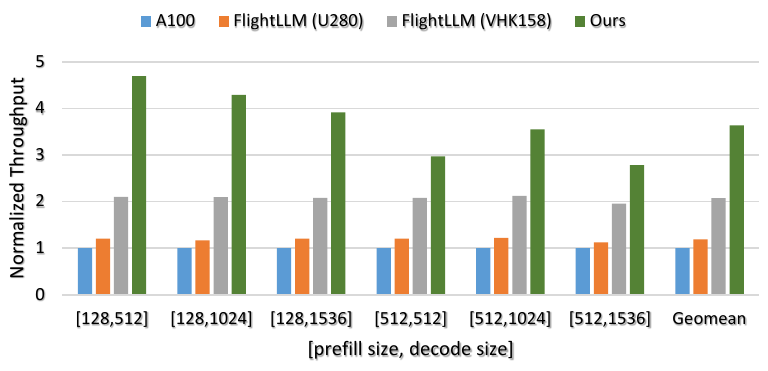}
\caption{Normalized throughput of AccLLM, FlightLLM, and A100 GPU.}
\label{fig:normalized_latency}\vspace{-1.6em}
\end{figure}

\subsection{Performance Evaluation of AccLLM Accelerator} 
The performance metrics of deploying LLMs on different hardware platforms are presented in Table \ref{tab:hw_accelerator}.
We can see that: {(1)} Compared to FP16 Llama-2-7B\cite{Touvron2023Llama2O} executed on an A100 GPU, our algorithm and hardware co-optimized AccLLM achieves \lyb{$\uparrow$$\mathbf{3.64}\times$} throughput and \lyb{$\uparrow$$\mathbf{24.8}\times$} energy efficiency. {(2)} When compared with W$4$A$16$ quantized and sparse ChatGLM2-6B \cite{Zeng2024ChatGLMAF} on its dedicated edge accelerator, EdgeLLM \cite{Huang2024EdgeLLMAH}, we offer \lyb{$\uparrow$$\mathbf{2.18}\times$} throughput and \lyb{$\uparrow$$\mathbf{3.37}\times$} energy efficiency. {(3)} Compared to the most competitive baseline, FlightLLM, which deploys mixed-quantized and sparse Llama-2-7B \cite{Touvron2023Llama2O} on two FPGA platforms \cite{Zeng2024FlightLLMEL} (Xilinx Versal VHK158 and Alveo U280), we provide \lyb{$\uparrow$$\mathbf{1.77}\times$} and \lyb{$\uparrow$$\mathbf{2.98}\times$} throughput, along with \lyb{$\uparrow$$\mathbf{8.27}\times$} and \lyb{$\uparrow$$\mathbf{4.07}\times$} energy efficiency, respectively. 
Our performance advantages over FlightLLM stem from two key innovations: (1) the W$2$A$8$KV$4$ quantization scheme that alleviates the bandwidth bottleneck during the decode stage and (2) the flexible DSP packing strategy that maximizes DSP utilization, together leading to significant improvements to both throughput and energy efficiency.

We further compare the normalized throughput with the A100 GPU and FlightLLM when processing different input prefill and output decode token sizes. As shown in Fig. \ref{fig:normalized_latency}, we observe that: 
{(1)} Benefiting from both algorithmic and hardware optimizations, our method outperforms the A100 GPU and FlightLLM (U280 and VHK158) across various input and output token lengths, achieving a \lyb{$\mathbf{1.77}\times$$\sim$$\mathbf{3.64}\times$} improvement in geometric mean throughput. 
{(2)} Particularly, when the input and output token sizes are $128$ and $512$, our approach demonstrates much better performance. 
This performance gain stems from our accelerator's reconfigurability, which efficiently accommodates both prefill and decode stages.

\begin{figure}[!t]
\centering
\setlength{\abovecaptionskip}{-0.1cm}
\includegraphics[width=1\columnwidth]{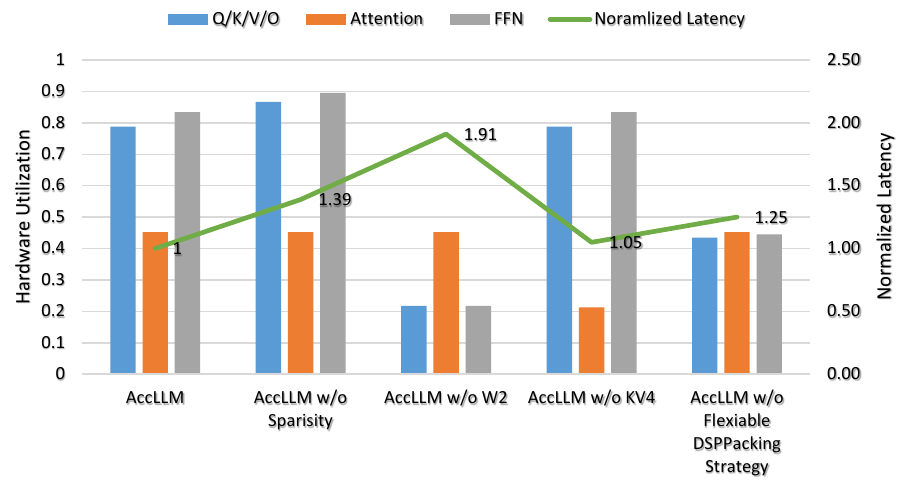}
\caption{The ablation studies on the end-to-end latency and hardware utilization of AccLLM evaluated on Llama-2-7B \cite{Touvron2023Llama2O} containing Q/K/V/O, Attention, and FFN in decode stage.}
\label{fig:speedup}\vspace{-1em}
\end{figure}


\begin{figure}[!t]
\centering
\setlength{\abovecaptionskip}{-0.1cm}
\includegraphics[width=1\columnwidth]{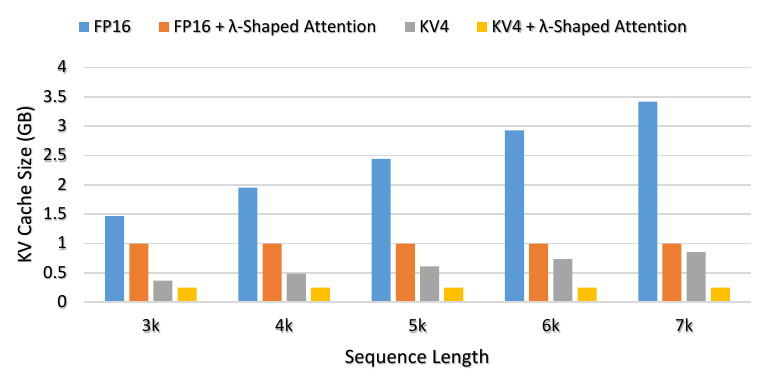}
\caption{The ablation studies on KV cache size of AccLLM.}
\label{fig:kvcache}\vspace{-1.7em}
\end{figure}

\textbf{Hardware Ablation Studies:} We further conduct ablation studies on latency reduction and hardware utilization of different methods used in AccLLM. As shown in Fig. \ref{fig:speedup}, we observe the following: {(1)} Although $2$:$4$ semi-structured pruning provides limited improvements in hardware utilization, it effectively eliminates redundant parameters, thereby enhancing throughput and achieving \lyb{$\mathbf{1.39}\times$} speedup. {(2)} The W$2$ quantization significantly reduces bandwidth requirements in linear layers and improves hardware utilization (approximately \lyb{$\mathbf{4}\times$} compared to conventional W$8$ quantization), leading to \lyb{$\mathbf{1.91}\times$} speedup. {(3)} The KV$4$ quantization alleviates bandwidth demands and enhances hardware utilization by \lyb{$\mathbf{2}\times$} during attention computation, resulting in a \lyb{$\mathbf{1.05}\times$} speedup, primarily since attention computations account for only a small fraction of total computations (as shown in Fig. \ref{fig:historgram}). Despite the modest speedup, the KV$4$ reduces memory requirements of KV cache by $\mathbf{75}\%$ compared to the FP16 counterpart, which will be discussed in the following paragraph. {(4)} The flexible DSP packing strategy optimizes DSP utilization, achieving approximately \lyb{$\mathbf{2}\times$} improvement in linear layers, contributing to a \lyb{$\mathbf{1.28}\times$} overall speedup. 

We also evaluate the memory footprint reductions of different attention optimizations related to the critical KV cache on varying sequence lengths. As demonstrated in Fig. \ref{fig:kvcache}: {(1)} The $\Lambda$-shaped attention effectively limits KV cache size to a fixed $\mathbf{1}$ GB (equivalent to 2$+$2044 selected tokens), regardless of the input sequence length. {(2)} KV$4$ quantization reduces KV cache size by \lyb{$\mathbf{75}\%$} compared to the FP16 baseline. {(3)} The combination of $\Lambda$-shaped attention and KV$4$ quantization achieves a remarkable reduction in KV cache size, limiting it to just $\mathbf{0.25}$ GB, highlighting the effectiveness of our approach in minimizing the memory footprint of KV cache.

\section{Conclusion}
\label{sec:conclusion}
In this paper, we have proposed, developed, and validated AccLLM, a comprehensive algorithm-hardware co-design framework that enables efficient and fast inference for LLMs on the FPGA platform. Specifically, at the algorithmic level, we proposed an aggressive compression algorithm, which combines $2$:$4$ semi-structured pruning, an innovative W2A8KV4 quantization scheme, and $\Lambda$-shaped attention, thus enhancing computational efficiency, reducing memory and bandwidth overhead, and enabling efficient long-sequence generation.
At the hardware level, we design an FPGA-based dedicated accelerator that features a reconfigurable computing engine to fully unleash our algorithmic benefits and boost hardware efficiency. Extensive experimental results consistently demonstrate our effectiveness, achieving up to \lyb{$\uparrow$$\mathbf{4.07}\times$} energy efficiency and \lyb{$\uparrow$$\mathbf{2.98}\times$} throughput compared to state-of-the-art LLM accelerators.

\bibliographystyle{IEEEtran}
\bibliography{main}

\begin{thebibliography}{10}
\providecommand{\url}[1]{#1}
\csname url@samestyle\endcsname
\providecommand{\newblock}{\relax}
\providecommand{\bibinfo}[2]{#2}
\providecommand{\BIBentrySTDinterwordspacing}{\spaceskip=0pt\relax}
\providecommand{\BIBentryALTinterwordstretchfactor}{4}
\providecommand{\BIBentryALTinterwordspacing}{\spaceskip=\fontdimen2\font plus
\BIBentryALTinterwordstretchfactor\fontdimen3\font minus \fontdimen4\font\relax}
\providecommand{\BIBforeignlanguage}[2]{{%
\expandafter\ifx\csname l@#1\endcsname\relax
\typeout{** WARNING: IEEEtran.bst: No hyphenation pattern has been}%
\typeout{** loaded for the language `#1'. Using the pattern for}%
\typeout{** the default language instead.}%
\else
\language=\csname l@#1\endcsname
\fi
#2}}
\providecommand{\BIBdecl}{\relax}
\BIBdecl

\bibitem{Touvron2023LLaMAOA}
H.~Touvron \emph{et~al.}, ``Llama: Open and efficient foundation language models,'' \emph{ArXiv}, vol. abs/2302.13971, 2023.

\bibitem{Achiam2023GPT4TR}
O.~J. Achiam \emph{et~al.}, ``Gpt-4 technical report,'' 2023.

\bibitem{Zhang2022OPTOP}
S.~Zhang \emph{et~al.}, ``Opt: Open pre-trained transformer language models,'' \emph{ArXiv}, vol. abs/2205.01068, 2022.

\bibitem{Touvron2023Llama2O}
H.~Touvron \emph{et~al.}, ``Llama 2: Open foundation and fine-tuned chat models,'' \emph{ArXiv}, vol. abs/2307.09288, 2023.

\bibitem{Naveed2023ACO}
H.~Naveed, A.~U. Khan, S.~Qiu, M.~Saqib, S.~Anwar, M.~Usman, N.~Barnes, and A.~S. Mian, ``A comprehensive overview of large language models,'' \emph{ArXiv}, vol. abs/2307.06435, 2023.

\bibitem{Chen2021EvaluatingLL}
M.~Chen \emph{et~al.}, ``Evaluating large language models trained on code,'' \emph{ArXiv}, vol. abs/2107.03374, 2021.

\bibitem{Zhang2023BenchmarkingLL}
T.~Zhang, F.~Ladhak, E.~Durmus, P.~Liang, K.~McKeown, and T.~Hashimoto, ``Benchmarking large language models for news summarization,'' \emph{Transactions of the Association for Computational Linguistics}, vol.~12, pp. 39--57, 2023.

\bibitem{Kamalloo2023EvaluatingOQ}
E.~Kamalloo, N.~Dziri, C.~L.~A. Clarke, and D.~Rafiei, ``Evaluating open-domain question answering in the era of large language models,'' \emph{ArXiv}, vol. abs/2305.06984, 2023.

\bibitem{Xu2024LlamaFAE}
H.~Xu, Y.~Li, and S.~Ji, ``Llamaf: An efficient llama2 architecture accelerator on embedded fpgas,'' \emph{ArXiv}, vol. abs/2409.11424, 2024.

\bibitem{Huang2024EdgeLLMAH}
M.~Huang, A.~Shen, K.~Li, H.~Peng, B.~Li, and H.~Yu, ``Edgellm: A highly efficient cpu-fpga heterogeneous edge accelerator for large language models,'' \emph{ArXiv}, vol. abs/2407.21325, 2024.

\bibitem{Friha2024LLMBasedEI}
O.~Friha, M.~A. Ferrag, B.~Kantarci, B.~Cakmak, A.~Ozgun, and N.~Ghoualmi-Zine, ``Llm-based edge intelligence: A comprehensive survey on architectures, applications, security and trustworthiness,'' \emph{IEEE Open Journal of the Communications Society}, vol.~5, pp. 5799--5856, 2024.

\bibitem{Aminabadi2022DeepSpeedIE}
R.~Y. Aminabadi \emph{et~al.}, ``Deepspeed- inference: Enabling efficient inference of transformer models at unprecedented scale,'' \emph{SC22: International Conference for High Performance Computing, Networking, Storage and Analysis}, pp. 1--15, 2022.

\bibitem{Dao2022FlashAttentionFA}
T.~Dao, D.~Y. Fu, S.~Ermon, A.~Rudra, and C.~R'e, ``Flashattention: Fast and memory-efficient exact attention with io-awareness,'' \emph{ArXiv}, vol. abs/2205.14135, 2022.

\bibitem{Kim2023FullSO}
S.~Kim, C.~Hooper, T.~Wattanawong, M.~Kang, R.~Yan, H.~Genç, G.~Dinh, Q.~Huang, K.~Keutzer, M.~W. Mahoney, Y.~S. Shao, and A.~Gholami, ``Full stack optimization of transformer inference: a survey,'' \emph{ArXiv}, vol. abs/2302.14017, 2023.

\bibitem{Pope2022EfficientlyST}
R.~Pope, S.~Douglas, A.~Chowdhery, J.~Devlin, J.~Bradbury, A.~Levskaya, J.~Heek, K.~Xiao, S.~Agrawal, and J.~Dean, ``Efficiently scaling transformer inference,'' \emph{ArXiv}, vol. abs/2211.05102, 2022.

\bibitem{Zeng2024FlightLLMEL}
S.~Zeng \emph{et~al.}, ``Flightllm: Efficient large language model inference with a complete mapping flow on fpgas,'' \emph{Proceedings of the 2024 ACM/SIGDA International Symposium on Field Programmable Gate Arrays}, 2024.

\bibitem{Frantar2023SparseGPTML}
E.~Frantar and D.~Alistarh, ``Sparsegpt: Massive language models can be accurately pruned in one-shot,'' \emph{ArXiv}, vol. abs/2301.00774, 2023.

\bibitem{Sun2023ASA}
M.~Sun, Z.~Liu, A.~Bair, and J.~Z. Kolter, ``A simple and effective pruning approach for large language models,'' \emph{ArXiv}, vol. abs/2306.11695, 2023.

\bibitem{Lin2024QServeWQ}
Y.~Lin, H.~Tang, S.~Yang, Z.~Zhang, G.~Xiao, C.~Gan, and S.~Han, ``Qserve: W4a8kv4 quantization and system co-design for efficient llm serving,'' \emph{ArXiv}, vol. abs/2405.04532, 2024.

\bibitem{Xiao2022SmoothQuantAA}
G.~Xiao, J.~Lin, M.~Seznec, J.~Demouth, and S.~Han, ``Smoothquant: Accurate and efficient post-training quantization for large language models,'' \emph{ArXiv}, vol. abs/2211.10438, 2022.

\bibitem{lin2024awq}
J.~Lin, J.~Tang, H.~Tang, S.~Yang, W.-M. Chen, W.-C. Wang, G.~Xiao, X.~Dang, C.~Gan, and S.~Han, ``Awq: Activation-aware weight quantization for on-device llm compression and acceleration,'' \emph{Proceedings of Machine Learning and Systems}, vol.~6, pp. 87--100, 2024.

\bibitem{Agrawal2023SARATHIEL}
A.~Agrawal, A.~Panwar, J.~Mohan, N.~Kwatra, B.~S. Gulavani, and R.~Ramjee, ``Sarathi: Efficient llm inference by piggybacking decodes with chunked prefills,'' \emph{ArXiv}, vol. abs/2308.16369, 2023.

\bibitem{Zhao2024PrepackingAS}
S.~Zhao, D.~Israel, G.~V. den Broeck, and A.~Grover, ``Prepacking: A simple method for fast prefilling and increased throughput in large language models,'' \emph{ArXiv}, vol. abs/2404.09529, 2024.

\bibitem{Ge2023ModelTY}
S.~Ge, Y.~Zhang, L.~Liu, M.~Zhang, J.~Han, and J.~Gao, ``Model tells you what to discard: Adaptive kv cache compression for llms,'' \emph{ArXiv}, vol. abs/2310.01801, 2023.

\bibitem{Xiao2023EfficientSL}
G.~Xiao, Y.~Tian, B.~Chen, S.~Han, and M.~Lewis, ``Efficient streaming language models with attention sinks,'' \emph{ArXiv}, vol. abs/2309.17453, 2023.

\bibitem{Kwon2023EfficientMM}
W.~Kwon, Z.~Li, S.~Zhuang, Y.~Sheng, L.~Zheng, C.~H. Yu, J.~E. Gonzalez, H.~Zhang, and I.~Stoica, ``Efficient memory management for large language model serving with pagedattention,'' \emph{Proceedings of the 29th Symposium on Operating Systems Principles}, 2023.

\bibitem{Qin2023FACTFC}
Y.~Qin, Y.~Wang, D.~Deng, Z.~Zhao, X.~Yang, L.~Liu, S.~Wei, Y.~Hu, and S.~Yin, ``Fact: Ffn-attention co-optimized transformer architecture with eager correlation prediction,'' \emph{Proceedings of the 50th Annual International Symposium on Computer Architecture}, 2023.

\bibitem{Ham2021ELSAHC}
T.~J. Ham, Y.~Lee, S.~H. Seo, S.-U. Kim, H.~Choi, S.~Jung, and J.~W. Lee, ``Elsa: Hardware-software co-design for efficient, lightweight self-attention mechanism in neural networks,'' \emph{2021 ACM/IEEE 48th Annual International Symposium on Computer Architecture (ISCA)}, pp. 692--705, 2021.

\bibitem{Lu2021SangerAC}
L.~Lu, Y.~Jin, H.~Bi, Z.~Luo, P.~Li, T.~Wang, and Y.~Liang, ``Sanger: A co-design framework for enabling sparse attention using reconfigurable architecture,'' \emph{MICRO-54: 54th Annual IEEE/ACM International Symposium on Microarchitecture}, 2021.

\bibitem{Press2021TrainST}
O.~Press, N.~A. Smith, and M.~Lewis, ``Train short, test long: Attention with linear biases enables input length extrapolation,'' \emph{ArXiv}, vol. abs/2108.12409, 2021.

\bibitem{han2024lm}
C.~Han, Q.~Wang, H.~Peng, W.~Xiong, Y.~Chen, H.~Ji, and S.~Wang, ``Lm-infinite: Zero-shot extreme length generalization for large language models,'' in \emph{Proceedings of the 2024 Conference of the North American Chapter of the Association for Computational Linguistics: Human Language Technologies (Volume 1: Long Papers)}, 2024, pp. 3991--4008.

\bibitem{Shao2023OneShotSM}
H.~Shao, B.~Liu, and Y.~Qian, ``One-shot sensitivity-aware mixed sparsity pruning for large language models,'' \emph{ICASSP 2024 - 2024 IEEE International Conference on Acoustics, Speech and Signal Processing (ICASSP)}, pp. 11\,296--11\,300, 2023.

\bibitem{Syed2023PruneAT}
A.~Syed, P.~H. Guo, and V.~Sundarapandiyan, ``Prune and tune: Improving efficient pruning techniques for massive language models,'' in \emph{Tiny Papers @ ICLR}, 2023.

\bibitem{Xu2024BESAPL}
P.~Xu, W.~Shao, M.~Chen, S.~Tang, K.-C. Zhang, P.~Gao, F.~An, Y.~Qiao, and P.~Luo, ``Besa: Pruning large language models with blockwise parameter-efficient sparsity allocation,'' \emph{ArXiv}, vol. abs/2402.16880, 2024.

\bibitem{Ma2023LLMPrunerOT}
X.~Ma, G.~Fang, and X.~Wang, ``Llm-pruner: On the structural pruning of large language models,'' \emph{ArXiv}, vol. abs/2305.11627, 2023.

\bibitem{Liu2023DejaVC}
Z.~Liu, J.~Wang, T.~Dao, T.~Zhou, B.~Yuan, Z.~Song, A.~Shrivastava, C.~Zhang, Y.~Tian, C.~R{\'e}, and B.~Chen, ``Deja vu: Contextual sparsity for efficient llms at inference time,'' in \emph{International Conference on Machine Learning}, 2023.

\bibitem{Li2023LoSparseSC}
Y.~Li, Y.~Yu, Q.~Zhang, C.~Liang, P.~He, W.~Chen, and T.~Zhao, ``Losparse: Structured compression of large language models based on low-rank and sparse approximation,'' in \emph{International Conference on Machine Learning}, 2023.

\bibitem{Molchanov2016PruningCN}
P.~Molchanov, S.~Tyree, T.~Karras, T.~Aila, and J.~Kautz, ``Pruning convolutional neural networks for resource efficient transfer learning,'' \emph{ArXiv}, vol. abs/1611.06440, 2016.

\bibitem{He2017ChannelPF}
Y.~He, X.~Zhang, and J.~Sun, ``Channel pruning for accelerating very deep neural networks,'' \emph{2017 IEEE International Conference on Computer Vision (ICCV)}, pp. 1398--1406, 2017.

\bibitem{Mishra2021AcceleratingSD}
A.~K. Mishra, J.~A. Latorre, J.~Pool, D.~Stosic, D.~Stosic, G.~Venkatesh, C.~Yu, and P.~Micikevicius, ``Accelerating sparse deep neural networks,'' \emph{ArXiv}, vol. abs/2104.08378, 2021.

\bibitem{Min2021MetaICLLT}
S.~Min, M.~Lewis, L.~Zettlemoyer, and H.~Hajishirzi, ``Metaicl: Learning to learn in context,'' \emph{ArXiv}, vol. abs/2110.15943, 2021.

\bibitem{Wei2021FinetunedLM}
J.~Wei, M.~Bosma, V.~Zhao, K.~Guu, A.~W. Yu, B.~Lester, N.~Du, A.~M. Dai, and Q.~V. Le, ``Finetuned language models are zero-shot learners,'' \emph{ArXiv}, vol. abs/2109.01652, 2021.

\bibitem{Liu2022FewShotPF}
H.~Liu, D.~Tam, M.~Muqeeth, J.~Mohta, T.~Huang, M.~Bansal, and C.~Raffel, ``Few-shot parameter-efficient fine-tuning is better and cheaper than in-context learning,'' \emph{ArXiv}, vol. abs/2205.05638, 2022.

\bibitem{Frantar2022GPTQAP}
E.~Frantar, S.~Ashkboos, T.~Hoefler, and D.~Alistarh, ``Gptq: Accurate post-training quantization for generative pre-trained transformers,'' \emph{ArXiv}, vol. abs/2210.17323, 2022.

\bibitem{Hong2022DFXAL}
S.~Hong, S.~Moon, J.~Kim, S.~Lee, M.~Kim, D.~Lee, and J.-Y. Kim, ``Dfx: A low-latency multi-fpga appliance for accelerating transformer-based text generation,'' \emph{2022 55th IEEE/ACM International Symposium on Microarchitecture (MICRO)}, pp. 616--630, 2022.

\bibitem{Radford2019LanguageMA}
A.~Radford, J.~Wu, R.~Child, D.~Luan, D.~Amodei, and I.~Sutskever, ``Language models are unsupervised multitask learners,'' 2019.

\bibitem{Zhang2024TinyLlamaAO}
P.~Zhang, G.~Zeng, T.~Wang, and W.~Lu, ``Tinyllama: An open-source small language model,'' \emph{ArXiv}, vol. abs/2401.02385, 2024.

\bibitem{Tang2024QuestQS}
J.~Tang, Y.~Zhao, K.~Zhu, G.~Xiao, B.~Kasikci, and S.~Han, ``Quest: Query-aware sparsity for efficient long-context llm inference,'' \emph{ArXiv}, vol. abs/2406.10774, 2024.

\bibitem{Kao2021FLATAO}
S.-C. Kao, S.~Subramanian, G.~Agrawal, A.~Yazdanbakhsh, and T.~Krishna, ``Flat: An optimized dataflow for mitigating attention bottlenecks,'' \emph{Proceedings of the 28th ACM International Conference on Architectural Support for Programming Languages and Operating Systems, Volume 2}, 2021.

\bibitem{Williams2008RooflineAI}
S.~Williams, A.~Waterman, and D.~Patterson, ``Roofline: an insightful visual performance model for multicore architectures,'' \emph{Commun. ACM}, vol.~52, no.~4, p. 65–76, apr 2009.

\bibitem{Merity2016PointerSM}
S.~Merity, C.~Xiong, J.~Bradbury, and R.~Socher, ``Pointer sentinel mixture models,'' \emph{ArXiv}, vol. abs/1609.07843, 2016.

\bibitem{Hassibi1993OptimalBS}
B.~Hassibi, D.~G. Stork, and G.~J. Wolff, ``Optimal brain surgeon and general network pruning,'' \emph{IEEE International Conference on Neural Networks}, pp. 293--299 vol.1, 1993.

\bibitem{Shao2023OmniQuantOC}
W.~Shao, M.~Chen, Z.~Zhang, P.~Xu, L.~Zhao, Z.~Li, K.~Zhang, P.~Gao, Y.~J. Qiao, and P.~Luo, ``Omniquant: Omnidirectionally calibrated quantization for large language models,'' \emph{ArXiv}, vol. abs/2308.13137, 2023.

\bibitem{Liao2024ApiQFO}
B.~Liao and C.~Monz, ``Apiq: Finetuning of 2-bit quantized large language model,'' in \emph{Conference on Empirical Methods in Natural Language Processing}, 2024.

\bibitem{Liu2021NonuniformtoUniformQT}
Z.~Liu, K.-T. Cheng, D.~Huang, E.~P. Xing, and Z.~Shen, ``Nonuniform-to-uniform quantization: Towards accurate quantization via generalized straight-through estimation,'' \emph{2022 IEEE/CVF Conference on Computer Vision and Pattern Recognition (CVPR)}, pp. 4932--4942, 2021.

\bibitem{Hu2021LoRALA}
J.~E. Hu, Y.~Shen, P.~Wallis, Z.~Allen-Zhu, Y.~Li, S.~Wang, and W.~Chen, ``Lora: Low-rank adaptation of large language models,'' \emph{ArXiv}, vol. abs/2106.09685, 2021.

\bibitem{Dettmers2023QLoRAEF}
T.~Dettmers, A.~Pagnoni, A.~Holtzman, and L.~Zettlemoyer, ``Qlora: Efficient finetuning of quantized llms,'' \emph{ArXiv}, vol. abs/2305.14314, 2023.

\bibitem{Bai2024SWATSA}
Z.~Bai, P.~Dangi, H.~Li, and T.~Mitra, ``Swat: Scalable and efficient window attention-based transformers acceleration on fpgas,'' \emph{ArXiv}, vol. abs/2405.17025, 2024.

\bibitem{Raffel2019ExploringTL}
C.~Raffel, N.~M. Shazeer, A.~Roberts, K.~Lee, S.~Narang, M.~Matena, Y.~Zhou, W.~Li, and P.~J. Liu, ``Exploring the limits of transfer learning with a unified text-to-text transformer,'' \emph{J. Mach. Learn. Res.}, vol.~21, pp. 140:1--140:67, 2019.

\bibitem{Dass2022ViTALiTyUL}
J.~Dass, S.~Wu, H.~Shi, C.~Li, Z.~Ye, Z.~Wang, and Y.~Lin, ``Vitality: Unifying low-rank and sparse approximation for vision transformer acceleration with a linear taylor attention,'' \emph{2023 IEEE International Symposium on High-Performance Computer Architecture (HPCA)}, pp. 415--428, 2022.

\bibitem{Shi2024TrioViTPQ}
H.~Shi, H.~Shao, W.~Mao, and Z.~Wang, ``Trio-vit: Post-training quantization and acceleration for softmax-free efficient vision transformer,'' \emph{IEEE Transactions on Circuits and Systems I: Regular Papers}, vol.~72, pp. 1296--1307, 2024.

\bibitem{Zeng2024ChatGLMAF}
T.~G.~A. Zeng \emph{et~al.}, ``Chatglm: A family of large language models from glm-130b to glm-4 all tools,'' \emph{ArXiv}, vol. abs/2406.12793, 2024.

\end{thebibliography}


\end{document}